# KINEMATIC PROPERTIES OF THE PAULI EQUATION

**E.E. Perepelkin[a,c,d,*], B.I. Sadovnikov[a], N.G. Inozemtseva[b], V.A. Svetovidov[a]**

[a] *Faculty of Physics, Lomonosov Moscow State University, Moscow, 119991 Russia*
[b] *Moscow Technical University of Communications and Informatics, Moscow, 123423 Russia*
[c] *Dubna State University, Moscow region, Dubna,141980 Russia*
[d] *Joint Institute for Nuclear Research, Moscow region, Dubna,141980 Russia*
**Corresponding author: pevgeny@jinr.ru*

## Abstract

Based on the Wigner-Vlasov formalism, this paper investigates the kinematic properties of the Pauli equation. It is shown that the probability current associated with the Pauli equation can be represented as a superposition of two currents with certain expansion coefficients. Each of these currents corresponds to a particular component of the spinor. The expansion coefficients effectively serve as weighting functions that determine the probability contribution of the corresponding spinor component. Therefore, each spin projection corresponds to its own probability flux. A new system of the Hamilton-Jacobi equations and also a system of motion equations in electromagnetic fields are obtained, taking into account the interaction between the spin and the magnetic field. To illustrate how these equations can be applied we have investigated the quantum system kinematics in detail using an exact solution of the Pauli equation in the presence of a uniform magnetic field and an asymmetric quadratic potential.

## Introduction

Quantum mechanics is one of the most nontrivial branches of physics and currently possesses up to nine distinct formulations [1]. For many years, attempts to understand or interpret quantum-mechanical phenomena from the standpoint of classical physics encountered significant difficulties [2-5]. A new opportunity emerged with the development of quantum mechanics in phase space through the pioneering works of Weyl and Wigner [6, 7]. Despite the Heisenberg uncertainty principle, the existence of a quasi-probability distribution function (the Wigner function) in phase space has proven to be highly fruitful for a wide range of applied problems [8-12]. The mathematical formalism based on the Wigner function makes it possible to approach the understanding of quantum processes from a classical-mechanical perspective. Using the Wigner-Vlasov formalism, one can establish a mathematically rigorous correspondence between quantum and classical systems [13-15]. At the heart of the Wigner-Vlasov formalism lies an infinite self-linked chain of the Vlasov equations for the distribution functions $f_n$, constructed from the fundamental principle – the law of probability conservation in a generalized phase space $\Omega_\infty$ [16-17]

$$\hat{\pi}_n S_n = -Q_n,\ n\in\mathbb{N}, \tag{i.1}$$

$$\hat{\pi}_n \stackrel{\text{def}}{=} \partial_t + \vec{v}\cdot\nabla_r + \dot{\vec{v}}\cdot\nabla_v + \ldots + \left\langle \vec{\xi}^{\,n+1}\right\rangle\cdot\nabla_{\xi^n}, \qquad S_n \stackrel{\text{def}}{=} \operatorname{Ln} f_n, \qquad Q_n \stackrel{\text{def}}{=} \operatorname{div}_{\xi^n}\left\langle \vec{\xi}^{\,n+1}\right\rangle,$$

$$f_n \stackrel{\text{def}}{=} \prod_{s=n+1}^{+\infty}\int_{\mathbb{R}^3} f_\infty\left(\vec{\xi},t\right)d^3\xi^s, \qquad \left\langle \vec{\xi}^{\,n+1}\right\rangle \stackrel{\text{def}}{=} \frac{1}{f_n}\int_{\mathbb{R}^3} f_\infty\left(\vec{\xi},t\right)\vec{\xi}^{\,n+1}d^3\xi^{n+1}\prod_{s=n+2}^{+\infty}\int_{\mathbb{R}^3} d^3\xi^s,$$

where $\vec{\xi}^n = \left\{\vec{r},\vec{v},\dot{\vec{v}},...,\vec{r}^{(n)}\right\}^T \in \Omega_\infty$. The quantity $Q_n$ represents the density of dissipation sources of kinematic quantities, and $\hat{\pi}_n$ denotes the total time derivative along the phase trajectory in the $n$- dimensional phase subspace. According to (i.1) when $Q_n = 0$ one obtains an analogue of Liouville's theorem, since the probability density $f_n = const$ along the generalized phase trajectory.

Since the Vlasov chain of equations (i.1) is self-linked, it is necessary to cut it off at some equation of order $n$, corresponding to the required amount of kinematic information about the physical system. In the literature there is a well-known method of cutting the Vlasov chain off at the second equation ($n = 2$), by introducing the simplest dynamic approximation for the average field of accelerations $m\left\langle\dot{\vec{v}}\right\rangle = \vec{F}$, where $\vec{F}$ is the force. The second Vlasov equation for the function $f_2\left(\vec{r},\vec{v},t\right)$ is used in plasma physics, astrophysics, statistical physics, solid-state physics and accelerator physics [18-23]. In [24, 25] the dynamic Vlasov-Moyal approximation was proposed for $\left\langle\dot{\vec{v}}\right\rangle$

$$f_2\left\langle\dot{v}_k\right\rangle = \sum_{l=0}^{+\infty}\frac{(-1)^{l+1}(\hbar/2)^{2l}}{m^{2l+1}(2l+1)!}\frac{\partial U}{\partial x^k}\left(\vec{\nabla}_r\cdot\vec{\nabla}_v\right)^{2l} f_2, \tag{i.2}$$

which allows us to reduce Vlasov's second equation (i.1) to Moyal's evolution equation [26] for the Wigner function $W\left(\vec{r},\vec{p},t\right) = m^3 f_2\left(\vec{r},\vec{v},t\right)$. Let us notice that when deriving the Vlasov equations chain (i.1) the positivity property is not imposed on the function $f_n$. Consequently, negative values [27-32] of the Wigner function $W$ don't pose a problem. In the classical limit as $\hbar \ll 1$ the approximation (i.2) reduces to the ordinary Vlasov approximation $m\left\langle\dot{\vec{v}}\right\rangle = -\nabla_r U$. A similar result can be reached by averaging (i.2) over the velocity space $m\left\langle\left\langle\dot{\vec{v}}\right\rangle\right\rangle = -\nabla_r U$. The electromagnetic case of the Vlasov-Moyal approximation is considered in [33].

In [34, 36] for a non-negative density function $f_1\left(\vec{r},t\right) = |\Psi|^2 \geq 0$, $\Psi \in \mathbb{C}$ based on the Helmholtz decomposition [37] for the average probability flux velocity $\left\langle\vec{v}\right\rangle$

$$\left\langle\vec{v}\right\rangle\left(\vec{r},t\right) = -\alpha\nabla_r\Phi\left(\vec{r},t\right) + \gamma\vec{\mathrm{A}}_\Psi\left(\vec{r},t\right),\ \mathrm{div}_r\,\vec{\mathrm{A}}_\Psi = \chi, \tag{i.3}$$
$$\frac{\Psi}{\Psi^*} = e^{i2\varphi} \Rightarrow \mathrm{Arg}\left[\frac{\Psi}{\Psi^*}\right] = 2\varphi\left(\vec{r},t\right) + 2\pi k \stackrel{\mathrm{def}}{=} \Phi\left(\vec{r},t\right) = -i\,\mathrm{Ln}\left[\frac{\Psi}{\Psi^*}\right],\ k \in \mathbb{Z},$$

an analogue of the Schrödinger equation was derived from the first ($n = 1$) Vlasov equation (i.1)

$$\frac{i}{\beta}\frac{\partial\Psi}{\partial t} = -\alpha\beta\left(\hat{\mathrm{p}} - \frac{\gamma}{2\alpha\beta}\vec{\mathrm{A}}_\Psi\right)^2\Psi + U_S\Psi, \tag{i.4}$$

where $\alpha,\beta,\gamma$ are some constant values, $\hat{\mathrm{p}} = -\frac{i}{\beta}\nabla_r$ is some differential operator, and $U_S\left(\vec{r},t\right) \in \mathbb{R}$. Using the Euler representation for complex function $\Psi = |\Psi|\exp\left(i\varphi\right)$ and taking the first Vlasov equation into account, equation (i.4) takes the following form

$$-\frac{1}{\beta}\frac{\partial\varphi}{\partial t}=-\frac{1}{4\alpha\beta}\left|\langle\vec{v}\rangle\right|^2+\mathrm{V}_\Psi\overset{\text{def}}{=}\mathrm{H}, \tag{i.5}$$

$$\mathrm{V}_\Psi=U_S+\mathrm{Q}_S,\quad \mathrm{Q}_S=\frac{\alpha}{\beta}\frac{\Delta_r|\Psi|}{|\Psi|}, \tag{i.6}$$

which after the Helmholtz decomposition (i.3) turns into the following equation

$$\hat{\pi}_1\langle\vec{v}\rangle=\frac{d}{dt}\langle\vec{v}\rangle=-\gamma\left(\vec{\mathrm{E}}_\Psi+\langle\vec{v}\rangle\times\vec{\mathrm{B}}_\Psi\right), \tag{i.7}$$

$$\vec{\mathrm{E}}_\Psi\overset{\text{def}}{=}-\frac{\partial\vec{\mathrm{A}}_\Psi}{\partial t}-\frac{2\alpha\beta}{\gamma}\nabla_r\mathrm{V}_\Psi,\quad \vec{\mathrm{B}}_\Psi\overset{\text{def}}{=}\mathrm{curl}_r\vec{\mathrm{A}}_\Psi. \tag{i.8}$$

Substitution of values $\alpha=-\hbar/2m$, $\beta=1/\hbar$ and $\gamma=-q/m$ into (i.3)-(i.7) leads to the following results. Formally equation (i.4) turns into the electromagnetic form of the Schrödinger equation, where the vortex component $\vec{\mathrm{A}}_\Psi$ of probability flux velocity (i.3) plays the role of the vector potential of magnetic induction (i.8) $\vec{\mathrm{B}}_\Psi$. Real function $U_S$ corresponds to the potential energy. The equation (i.5) is the same as the Hamilton-Jacobi equation, which in addition to the potential energy $U_S$ contains the quantum potential $\mathrm{Q}_S$ (i.6) from the de Broglie–Bohm «wave-pilot» theory [38-40]. Equation (i.7) corresponds to the motion equation of the particle with charge $q$ in electromagnetic fields (i.8).

One should be accurate speaking about «electromagnetic» fields (i.8). In general case fields (i.8) are not required to satisfy Maxwell's equations. Being more precise, fields (i.8) definitely satisfy only two Maxwell's equations $\mathrm{div}_r\vec{\mathrm{B}}_\Psi=0$ and $-\partial_t\vec{\mathrm{B}}_\Psi=\mathrm{curl}_r\vec{\mathrm{E}}_\Psi$. The rest two Maxwell's equations can be satisfied when the system is self-consistent

$$\rho_\Psi\overset{\text{def}}{=}\varepsilon_0\,\mathrm{div}_r\vec{\mathrm{E}}_\Psi\overset{\text{def}}{=}\mathrm{div}_r\vec{\mathrm{D}}_\Psi=qf_1. \tag{i.9}$$

Usually in most of the studies only the influence of external fields on quantum system is taken into account, but the reverse effect of the quantum system itself on external fields is considered negligible. In such cases the physical system is not self-consistent and the condition (i.9) is violated. Nevertheless, there are exact solutions of non-linear problems in classical and quantum mechanics where this condition (i.9) is satisfied [13, 15]. In order to make the condition (i.9) weaker, the field from (i.8) can be rewritten into a sum of fields

$$\vec{\mathrm{A}}_\Psi=\vec{\mathrm{A}}+\vec{\mathrm{A}}_Q,\quad \vec{\mathrm{B}}_\Psi\overset{\text{def}}{=}\mathrm{curl}_r\vec{\mathrm{A}}_\Psi\overset{\text{def}}{=}\vec{\mathrm{B}}+\vec{\mathrm{B}}_Q, \tag{i.10}$$

$$\vec{\mathrm{E}}_\Psi=-\frac{\partial\vec{\mathrm{A}}}{\partial t}-\frac{1}{q}\nabla_r U_S-\frac{\partial\vec{\mathrm{A}}_Q}{\partial t}-\frac{1}{q}\nabla_r\mathrm{Q}_S=\vec{\mathrm{E}}_S+\vec{\mathrm{E}}_Q. \tag{i.11}$$

Superposition of (i.10)-(i.11) is natural since it contains classical potentials $\vec{\mathrm{A}}$ and $q\varphi_S=U_S$ satisfying Maxwell's equations as well as probability («quantum») potentials depending on wave function $\Psi$. The Lorenz $\Psi$-gauge condition [36] can be applied to the potentials (i.10)-(i.11).

$$\operatorname{div}_r \vec{\mathrm{A}}_\Psi + \frac{2\alpha\beta}{\gamma}\frac{1}{c^2}\frac{\partial \mathrm{V}_\Psi}{\partial t} = 0, \tag{i.12}$$

which can be in special case decomposed into a sum of two separate gauge conditions

$$\operatorname{div}_r \vec{\mathrm{A}} + \frac{1}{qc^2}\frac{\partial U_S}{\partial t} = 0, \qquad \operatorname{div}_r \vec{\mathrm{A}}_Q + \frac{1}{qc^2}\frac{\partial \mathrm{Q}_S}{\partial t} = 0. \tag{i.13}$$

The first condition in (i.13) is the well-known Lorenz gauge condition for electromagnetic potentials. The second condition in (i.13) is a gauge condition for quantum potentials. In general case the condition (i.12) can be satisfied, but if so then the conditions (i.13) may not be satisfied. In classical limit $\hbar \ll 1$ the expression for quantum potential $\mathrm{Q}_S = -\frac{\hbar^2}{2m}\frac{\Delta|\Psi|}{|\Psi|}$ can be negligible ($\mathrm{Q}_S \ll 1$), which leads to $\mathrm{V}_\Psi \approx U_S$ and transition of (i.5)-(i.8) to known classical equations.

If the condition of self-consistence (i.9) is satisfied, after having the (i.12) applied one should obtain the satisfaction of the Maxwell equations system

$$\operatorname{div}_r \vec{\mathrm{D}}_\Psi = \rho_\Psi, \;\; \operatorname{div}_r \vec{\mathrm{B}}_\Psi = 0, \;\; \operatorname{curl}_r \vec{\mathrm{E}}_\Psi = -\partial_t \vec{\mathrm{B}}_\Psi, \quad \partial_t \vec{\mathrm{D}}_\Psi + \vec{\mathrm{J}}_\Psi = \operatorname{curl}_r \vec{\mathrm{H}}_\Psi, \tag{i.14}$$

where $\vec{\mathrm{J}}_\Psi \overset{\text{def}}{=} \rho_\Psi \langle \vec{v} \rangle$, $\mu_0 \vec{\mathrm{H}}_\Psi \overset{\text{def}}{=} \vec{\mathrm{B}}_\Psi$. The quantum contribution corresponding to $\vec{\mathrm{B}}_Q$ and $\vec{\mathrm{E}}_Q$ can be partially «transferred» into the wave function, which should be «extended» for this reason by the transition from the scalar function $\Psi$ to spinor wave function $\psi \in \mathbb{C}^2$. According to this aim the expression [36] of the Helmholtz decomposition (i.3)

$$f_1 \langle \vec{v} \rangle = -i\alpha\left(\Psi^* \nabla_r \Psi + \Psi \nabla_r \Psi^*\right) + \gamma \vec{\mathrm{A}}_\Psi, \tag{i.15}$$

has been rewritten for spinor columns in terms of

$$\vec{J} = f_1 \langle \vec{v} \rangle = i\alpha\left(\psi^\dagger \mathrm{I}_2 \nabla_r \psi - \psi^T \mathrm{I}_2 \nabla_r \psi^*\right) + \gamma \psi^\dagger \vec{\mathrm{A}} \psi, \;\; f_1 = \psi^\dagger \psi = |\psi_1|^2 + |\psi_2|^2. \tag{i.16}$$

Substitution of the representation (i.16) into the first Vlasov equation (i.1) leads us to the analogue of the Pauli equation [35, 36]

$$\frac{i}{\beta}\mathrm{I}_2 \frac{\partial \psi}{\partial t} = -\alpha\beta\left[\vec{\sigma} \cdot \left(\hat{\mathrm{p}} - \frac{\gamma}{2\alpha\beta}\vec{\mathrm{A}}\right)\right]^2 \psi + \mathrm{I}_2 U \psi. \tag{i.17}$$

There are several questions and problems here. Firstly, since the Schrödinger and Pauli equations are non-relativistic, the decomposition (i.16), which uses derivatives only in coordinate space, should naturally be represented in the original form (i.3). Secondly, by analogy with (i.5) and (i.7), there should be the Hamilton-Jacobi equations and equations of motion for the Pauli equation (i.17). Is there a quantum potential for the Pauli equation, and what is its physical meaning? Since the Pauli equation takes spin into account, the corresponding motion equations must also take it into account. Thirdly, the solutions of the Pauli equation and the Schrödinger solutions can be used to obtain the solution of the first Vlasov equation. This means that there is a strict mathematical relationship between potentials and fields from the Schrödinger equation and Pauli. Fourthly, will the electromagnetic fields corresponding to the Schrödinger and Pauli equations satisfy Maxwell's equations, and under what conditions? Fifthly, it would be interesting to obtain an exact solution of the Pauli equation for a nontrivial case, which demonstrates the answers to the questions posed.

The purpose of this paper is to answer the above questions.

The work has the following structure. In section §1 a kinematic analysis of the expansion (i.16) is performed. As a result it is found that the vector field $\langle\vec{v}\rangle$ is representable as a superposition of fields $\langle\vec{v}_1\rangle$ and $\langle\vec{v}_2\rangle$ with expansion coefficients $\kappa_1$ и $\kappa_2$, that is $\langle\vec{v}\rangle=\kappa_1\langle\vec{v}_1\rangle+\kappa_2\langle\vec{v}_2\rangle$. Each flow $\langle\vec{v}_n\rangle$ is determined by the corresponding component $\psi_n$ of the spinor $\psi$ according to the Helmholtz decomposition (i.3). The physical meaning of the expansion coefficients $\kappa_n$ is the fraction of probability corresponding to the wave function $\psi_n$, that is $\kappa_n=|\psi_n|^2/|\psi|^2$. Similar relations are valid for the phases of the wave functions. An expression for the vector potential $\vec{\mathrm{A}}_Q$ (i.10) is explicitly obtained. For the Pauli equation, analogues of the Hamilton-Jacobi equations are obtained in the form of a system of two equations interconnected by the energy of the spin interaction. At the same time, each component $\psi_n$ corresponds to its quantum potential $\mathrm{Q}_n$. In addition to the potential $\mathrm{Q}_n$ the new equations include the energy $\mathrm{P}_n$ of the interaction of spin with the magnetic field. It is shown that the probability density components $|\psi_n|^2$ do not satisfy to continuity equations due to the nonzero right-hand side containing the spin interaction. But the sum of the equations for the densities $|\psi_n|^2$ gives the initial equation of continuity (the first Vlasov equation). Explicit expressions of the relationship between the wave functions $\psi$ and $\Psi$, as well as the scalar and vector potentials of the Pauli and Schrödinger equations are obtained. As a result, using the solutions of the Pauli equation, it is possible to «construct» another quantum system corresponding to it, described by the Schrödinger equation.

In section §2 analogues of the motion equations (i.7) are found for the Pauli equation in the form of a system of two differential equations connected by a spin interaction. That is the right-hand side of the motion equations, in addition to the Coulomb and Lorentz forces, contains a force acting on a magnetic dipole in an external magnetic field. Each equation of motion contains its own set of electromagnetic fields related to external electromagnetic fields. A clear connection has been established between the electromagnetic fields of the Schrödinger and Pauli equations. The conditions for the fulfillment of Maxwell's equations are considered. The self-consistency condition (i.9) for the Pauli equation and the possibility of violating calibration (i.13) when fulfilling (i.12) are investigated.

In section 3, an exact solution of the Pauli equation with a constant magnetic field and an asymmetric quadratic electric potential is constructed based on the Ψ-model [13]. The resulting solution demonstrates in detail all the results obtained in sections §1-2. A kinematic analysis of probability flows is carried out, and the system of Hamilton-Jacobi equations and motion equations is solved. The magnetic moment $\vec{M}_{s,\ell}$ of the quantum system is calculated, which is proportional to the Bohr magneton and consisting of two terms. The first term, $\mu_B\ell,\ \ell\in\mathbb{Z}$ is related to the vortex field of the $\Psi$-model, the second term is due to an external magnetic field. It is shown that the effect of the magnetic momentum $\vec{M}_{s,\ell}$ sign changing is possible with a certain ratio between the spin precession frequency $\omega_0$ and the frequency $\omega$, included in potential $U$. This effect is similar to magnetic resonance, but here it is caused not by a time-varying magnetic field, but by the frequency $\omega$ in the electric potential.

The conclusion contains general remarks on the results of the work. Appendices A, B provide proofs of the theorems and intermediate calculations.

## §1 The Hamilton-Jacobi equations

Let us start by analyzing the kinematic properties of the vector field of the probability flow and constructing the Hamilton-Jacobi equation for it. For the convenience of mathematical transformations, we will use the notation for the probability density function $f_1 = f$, and the index «$n$» $(n=1,2)$ will correspond to the spinor component.

**Lemma 1.** *Vector field of the probability flux $\langle \vec{v} \rangle(\vec{r},t)$ (i.3) is represented as a superposition of the fields $\langle \vec{v}_1 \rangle(\vec{r},t)$ and $\langle \vec{v}_2 \rangle(\vec{r},t)$ with expansion coefficients $\kappa_1, \kappa_2$*

$$\langle \vec{v} \rangle = \sum_{n=1}^{2} \kappa_n \langle \vec{v}_n \rangle = \frac{\hbar}{2m} \nabla_r \Phi - \frac{q}{m} \vec{\mathrm{A}}_\Psi, \qquad \langle \vec{v}_n \rangle = \frac{\hbar}{2m} \nabla_r \Phi_n - \frac{q}{m} \vec{\mathrm{A}}, \tag{1.1}$$

$$\Phi_n = -i \,\mathrm{Ln} \frac{\psi_n}{\psi_n^*} = 2\varphi_n + 2\pi k,\ k \in \mathbb{Z}, \quad \Phi = \sum_{n=1}^{2} \kappa_n \Phi_n, \quad f\kappa_n = |\psi_n|^2, \quad \sum_{n=1}^{2} \kappa_n = 1, \tag{1.2}$$

$$\vec{\mathrm{A}}_Q = \frac{\hbar}{2q} \sum_{n=1}^{2} \Phi_n \nabla_r \kappa_n, \quad \vec{\mathrm{B}}_Q = \frac{\hbar}{2q} \sum_{n=1}^{2} \nabla_r \Phi_n \times \nabla_r \kappa_n, \tag{1.3}$$

*where $\psi_n = |\psi_n| \exp(i\varphi_n)$, $|\psi_1|^2 + |\psi_2|^2 = f$, $n = 1,2$.*

Proof of lemma 1 is presented in Appendix A.

According to Lemma 1, the vector field of the probability flux $\langle \vec{v} \rangle$ (1.1) splits into two fluxes $\langle \vec{v}_1 \rangle$ and $\langle \vec{v}_2 \rangle$. Each of these fluxes is represented as the Helmholtz decomposition equation (1.1) with the same form of scalar potential (1.2). Note that expressions (1.2) are similar to expression (i.3). Lemma 1 implies that for each wave function $\psi_n$, $n=1,2$ corresponds to its potential probability flux $\frac{\hbar}{2m} \nabla_r \Phi_n$, but there is the same deterministic vortex flux $\vec{\mathrm{A}}$. From a physical point of view coefficients $\kappa_n$ (1.2) are weighting coefficients that determine which fraction $|\psi_n|^2 / f$ of the total density $f = |\psi_1|^2 + |\psi_2|^2$ goes to each «n»-flux $\langle \vec{v}_n \rangle$.

Despite the variability of expansion coefficients $\kappa_n$, the total flux $\langle \vec{v} \rangle$ has a scalar potential $\Phi$, which is a superposition (1.2) of scalar potentials $\Phi_n$. The transition to the indicated superposition of potentials is performed by separating the «probability/quantum» part of $\vec{\mathrm{A}}_Q$ (1.3) from the vortex field $\vec{\mathrm{A}}_\Psi$ in the form of decomposition $\vec{\mathrm{A}}_\Psi = \vec{\mathrm{A}} + \vec{\mathrm{A}}_Q$ (i.10).

Note that the obtained explicit expressions for the fields $\vec{\mathrm{A}}_Q$ and $\vec{\mathrm{B}}_Q$ (1.3) are fully expressed in terms of wave functions $\psi_n$, which once again emphasizes their probability/statistical nature. Unlike the external deterministic fields $\vec{\mathrm{A}}$ and $\vec{\mathrm{B}}$, fields $\vec{\mathrm{A}}_Q$ and $\vec{\mathrm{B}}_Q$ are eigenfields of a quantum system. This was discussed earlier in [36], but explicit expressions for the fields $\vec{\mathrm{A}}_Q$ and $\vec{\mathrm{B}}_Q$ are obtained in Lemma 1.

We emphasize once again that the transition from the scalar wave function $\Psi \in \mathbb{C}$ in the Schrödinger equation (i.4) to spinor one $\psi \in \mathbb{C}^2$ for the Pauli equation (i.17) leads to the splitting of the probability flux $\langle \vec{v} \rangle$ into two fluxes $\langle \vec{v}_1 \rangle$ and $\langle \vec{v}_2 \rangle$, and to selection of the probability/statistical components of the vortex field $\vec{\mathrm{A}}_Q$. As will be shown below, this fact is directly related to the concept of spin.

**Definition 1.** *Let us define the scalar product of vectors $\vec{a}$ u $\vec{b}$ in space with metrics ${}_n g$, $n = 1,2$ as*

$$\left(\vec{a} \cdot \vec{b}\right)_n \overset{\text{def}}{=} {}_n g_{sk} a^s b^k = {}_n a_k b^k = a^s \, {}_n b_s = {}_n g^{sk} \, {}_n a_s \, {}_n b_k, \tag{1.4}$$

$${}_n g \overset{\text{def}}{=} \operatorname{diag}\left(\sqrt{\kappa_n^{-1} - 1}, \sqrt{\kappa_n^{-1} - 1}, 1\right) \overset{\text{def}}{=} \left({}_n g_{sk}\right), \quad {}_2 g = \left({}_1 g\right)^{-1}, \tag{1.5}$$

*where by repeating indexes $s,k = 1...3$ it is implied the summation (Einstein's rule), and index $n$ is fixed.*

Metric (1.5) corresponds to an orthogonal curved coordinate system. The curvature of space is determined by the probability densities of $f_n = |\psi_n|^2$. For example, under $|\psi_1| = |\psi_2|$ the metric tensor ${}_n g = \mathrm{I}_3$, that is $\left(\vec{a} \cdot \vec{b}\right)_n = \vec{a} \cdot \vec{b}$. The property follows directly from the definition of ${}_n g$. Metric tensor $\left({}_1 g\right)^{-1}$ corresponds to the conjugate basis, therefore ${}_2 g$ is a metric in the conjugate space. Thus, from the point of geometry, the two-component spinor defines the metric of the space and the metric of the space conjugated to it.

**Theorem 1.** *The Pauli equation (i.17) can be transformed into two Hamilton-Jacobi equations*

$$-\hbar \partial_t \varphi_n = \frac{m}{2} \left|\langle \vec{v}_n \rangle\right|^2 + \mathcal{V}_n \overset{\text{def}}{=} H_n, \; \mathcal{V}_n = \mathrm{V}_n + \mathrm{P}_n, \quad \mathrm{V}_n = U + \mathrm{Q}_n, \tag{1.6}$$

$$\mathrm{Q}_n = -\frac{\hbar^2}{2m} \frac{\Delta_r |\psi_n|}{|\psi_n|}, \qquad \mathrm{P}_n = \left(\vec{\mu}_\pm \cdot \vec{\mathrm{B}}\right)_n = {}_n g_{sk} \mu_\pm^s \mathrm{B}^k, \quad n = 1,2, \tag{1.7}$$

$$\vec{\mu}_\pm\left(\varphi_{21}\right) \overset{\text{def}}{=} -\vec{\mu}_{\uparrow\downarrow} \overset{\text{def}}{=} -\frac{q\hbar}{2m}\left(\cos\varphi_{21} \;\; \sin\varphi_{21} \;\; \pm 1\right) \overset{\text{def}}{=} \left(\mu_\pm^s\right), \tag{1.8}$$

*where $\varphi_{21} = \varphi_2 - \varphi_1$. At the same time the continuity equation splits into the sum of two equations for probability densities $f_n = |\psi_n|^2$*

$$\partial_t f_n + \operatorname{div}_r \vec{J}_n = (-1)^n \frac{2}{\hbar} f_n \left(\vec{\mu}_a\left(\overline{\varphi}\right) \cdot \vec{\mathrm{B}}\right)_n, \tag{1.9}$$

*where $\vec{\mu}_a = \left(\vec{\mu}_+ + \vec{\mu}_-\right)/2$, $\overline{\varphi} = \varphi_{21} + \pi/2$ and $\vec{J}_n = f_n \langle \vec{v}_n \rangle$ is the density of n-th probability flux.*

Proof of the Theorem 1 is presented in Appendix A.

**Remark 1**. All values included in the formulation of Theorem 1, except the constant values $\hbar, m, q, n$, generally depend on the coordinate and time $(\vec{r}, t)$. This dependence is dropped from the formulation of the theorem 1 to make it brief. Let us denote that $\vec{\mu}_{\pm}$ is the same as $\vec{\mu}_{\uparrow\downarrow}$ and is the same as $\vec{\mu}_n$, where $\vec{\mu}_1 \equiv \vec{\mu}_{\uparrow} \equiv \vec{\mu}_+$ and $\vec{\mu}_2 \equiv \vec{\mu}_{\downarrow} \equiv \vec{\mu}_-$.

Note that equations (1.9), in contrast to the original continuity equation $\partial_t f + \mathrm{div}_r \left[ f \langle \vec{v} \rangle \right] = 0$, generally speaking, are not continuity equations because of the nonzero right-hand side. Nevertheless, summing up equations (1.9), we obtain the initial equation of continuity $\partial_t f + \mathrm{div}_r \vec{J} = 0$ with a current density $\vec{J} = \vec{J}_1 + \vec{J}_2 = f \langle \vec{v} \rangle$.

The vector $\vec{\mu}_a$ has only the transverse (radial) components and the magnetic field $\vec{\mathrm{B}} = \begin{pmatrix} 0 & 0 & \mathrm{B}^3 \end{pmatrix}$ has only the axial component. Therefore, the dot product $\left( \vec{\mu}_a \cdot \vec{\mathrm{B}} \right)_n = 0$. In this case the right-hand side in equations (1.9) is zero and the continuity equation is fulfilled for each flow $\vec{J}_n$.

On the left side of equations (1.6) there are analogues of partial time derivatives from the act $\mathrm{S}_n = \hbar \varphi_n$ of n-th flux. The right side of equations (1.6) is the sum of the kinetic energy $\frac{m}{2} \left| \langle \vec{v}_n \rangle \right|^2$ of n-th flux and the potential energy $\mathcal{V}_n$. Thus, the right-hand side of equation (1.6) corresponds to the Hamilton function $H_n$. The potential energy $\mathcal{V}_n$ differs from the energy $\mathrm{V}_\Psi$ (i.6) in the Schrödinger equation by the presence of an additional term $\mathrm{P}_n$. Absence of the term $\mathrm{P}_n$ would lead to the decomposition of the system of equations (1.6) into the two independent Hamilton-Jacobi equations. Each flux has its own potential energy $\mathrm{V}_n$ (1.6), which contains its own quantum potential $\mathrm{Q}_n$ (1.7). The potential energy $\mathrm{V}_n$ is an analogue of energy $\mathrm{V}_\Psi$ for the Schrödinger equation.

Let us consider in more detail the physical meaning of the energy $\mathrm{P}_n$, taking into account the self-consistent influence of the first flow on the second one and vice versa. The coefficient of the vector $\vec{\mu}_{\pm}$ at $q = -e$ ($e$ is elementary charge) equals to Bohr's magneton $\mu_B = e\hbar / 2m$, that is vector $\vec{\mu}_{\pm} = \mu_B \begin{pmatrix} \cos\varphi_{21} & \sin\varphi_{21} & \pm 1 \end{pmatrix}$ corresponds to the magnetic momentum. Note that the vector $\vec{\mu}_{\pm}$ is generally not constant since it depends on the phase difference $\varphi_{21}(\vec{r}, t)$. This dependence is present only in the radial directions, and the vertical (axial) component $\mu_{\pm}^{(z)} = \pm \mu_B$ is a constant value. In special case, when $|\psi_1| = |\psi_2|$ (${}_n g = \mathrm{I}_3$) the energy $\mathrm{P}_{\uparrow\downarrow} = -\vec{\mu}_{\uparrow\downarrow} \cdot \vec{\mathrm{B}}$ is a well-known expression for the energy of interaction between the magnetic dipole ($\vec{\mu}$) and external magnetic field $\vec{\mathrm{B}}$. In the general case, an additional factor arises in the form of the metric ${}_n g$ (1.7) when there is a curvature of space ${}_n g \neq \mathrm{I}_3$, caused by a violation of the balance between the first and second probability fluxes. Note that formally the form of expression (1.7)-(1.8) is similar to the expression for the intrinsic magnetic moment of an electron $\vec{\mu}_s = -g \mu_B \vec{\mathrm{s}}$ ($\vec{\mathrm{s}}$ is the spin of the particle) taking into account the $g$-factor, used in explaining the anomalous magnetic moment. The formal interpretation of such a «g-factor» from the standpoint of the expression ${}_n g_{sk} \mu_{\pm}^s$ (1.7) might be a metric of the space ${}_n g$, due to the violation of symmetry between the distributions of probability densities $|\psi_n|$, $n = 1, 2$ in space ($n = 1$) and conjugated space ($n = 2$).

Equations (1.6) can be written with respect to the phase difference $\varphi_{21}$, which is present in expressions (1.8)-(1.9). To do this, it is convenient to introduce the following notation.

**Definition 2.** *Let us define the column-vector* $\langle\mathcal{U}\rangle$ *and row-vector* $\langle\mathcal{U}\rangle^T$ *of average fluxes velocities* $\langle\vec{v}_n\rangle$, $n=1,2$, *and also the dot product of the pseudo-Euclidian metric* $\mathcal{G}$ *for these vectors*

$$\langle\mathcal{U}\rangle \overset{\text{def}}{=} \begin{bmatrix}\langle\vec{v}_1\rangle \\ \langle\vec{v}_2\rangle\end{bmatrix},\ \langle\mathcal{U}\rangle^T \overset{\text{def}}{=} \left[\langle\vec{v}_1\rangle \quad \langle\vec{v}_2\rangle\right],\ \mathcal{G} \overset{\text{def}}{=} \begin{pmatrix}-\mathrm{I}_3 & 0_3 \\ 0_3 & \mathrm{I}_3\end{pmatrix}, \tag{1.10}$$

$$\langle\mathcal{U}\rangle^2 \overset{\text{def}}{=} \langle\mathcal{U}\rangle^T \mathcal{G}\langle\mathcal{U}\rangle = \left|\langle\vec{v}_2\rangle\right|^2 - \left|\langle\vec{v}_1\rangle\right|^2, \tag{1.11}$$

*where the matrix* $\mathcal{G}$ *has dimension* $6\times6$, $\mathrm{I}_3$ *is the identity matrix, and* $0_3$ *is a zero matrix with dimension* $3\times3$.

Using (1.10)-(1.11) the difference between two equations (1.6) takes the form

$$-\hbar\partial_t\varphi_{21} = \frac{m}{2}\langle\mathcal{U}\rangle^2 + \mathcal{V}_{21}, \qquad \mathcal{V}_{21} \overset{\text{def}}{=} \mathrm{Q}_{21}+\mathrm{P}_{21} \overset{\text{def}}{=} \mathrm{Q}_2-\mathrm{Q}_1+\mathrm{P}_2-\mathrm{P}_1. \tag{1.12}$$

Note that there is no potential $U$ in equation (1.12), and it depends only on difference of $\mathrm{Q}_{12}$ quantum potentials, and the difference $\mathrm{P}_{12}$ of fluxes $\langle\vec{v}_n\rangle$ interaction energies. The term with $\langle\mathcal{U}\rangle^2$ in (1.12) formally corresponds to the «kinetic energy» with the caveat that it can be negative. Indeed, by virtue of the non-Euclidean metric $\mathcal{G}$, the dot product (1.11) does not have to be positive.

Let us consider the relationship between the Pauli equation and the Schrödinger equation. The idea is when we know the exact solution of one equation then we can construct an exact solution to another equation.

**Theorem 2.** *Let the spinor* $\psi(\vec{r},t)$ *is a solution of the Pauli equation (i.17) with potential energy* $U(\vec{r},t)$ *and vector potential* $\vec{\mathrm{A}}(\vec{r},t)$, *then the wave function*

$$\Psi(\vec{r},t) = \pm\sqrt{|\psi_1|^2+|\psi_2|^2}\exp(i\varphi), \qquad \varphi(\vec{r},t) = \sum_{n=1}^{2}\kappa_n\varphi_n + \pi l,\ l\in\mathbb{Z}, \tag{1.13}$$

*is a solution of the Schrödinger equation (i.4) with potential energy* $U_S(\vec{r},t)$, *vector potential* $\vec{\mathrm{A}}_\Psi(\vec{r},t)$ *and quantum potential* $\mathrm{Q}_S(\vec{r},t)$ *if these potentials satisfy the following relations*

$$\vec{\mathrm{A}}_\Psi = \vec{\mathrm{A}} - \frac{\hbar}{q}\varphi_{21}\nabla_r\kappa_1, \quad \mathrm{V}_\Psi = U_S + \mathrm{Q}_S = \mathcal{V} + \delta\mathcal{V}, \tag{1.14}$$

$$\mathcal{V} \overset{\text{def}}{=} \sum_{n=1}^{2}\kappa_n\mathcal{V}_n = U + \mathrm{Q} + \mathrm{P}, \qquad \mathrm{Q} = \sum_{n=1}^{2}\kappa_n\mathrm{Q}_n, \qquad \mathrm{P} = \sum_{n=1}^{2}\kappa_n\mathrm{P}_n, \tag{1.15}$$

$$\delta\mathcal{V} = \hbar\kappa_1\kappa_2\left(\varphi_{21}\partial_t K - \alpha\left|\nabla_r\varphi_{21}\right|^2\right), \qquad K \overset{\text{def}}{=} \ln(\kappa_1/\kappa_2), \tag{1.16}$$

$$\mathrm{Q}_S = \mathrm{Q} + \delta\mathrm{Q}, \qquad \delta\mathrm{Q} = -\frac{\hbar^2}{8m}\kappa_1\kappa_2\left|\nabla_r K\right|^2. \tag{1.17}$$

*In this case the Hamilton-Jacobi equation (i.5) is fulfilled for quantities* $\mathrm{V}_{\Psi}$ *and* $\langle\vec{v}\rangle$, *where the phase* $\varphi$ *of the wave function* $\Psi$ *has the form (1.13).*

Proof of the Theorem 2 is given in Appendix A.

It follows from Theorem 2 that the initial system of two Hamilton-Jacobi equations (1.6) for the Pauli equation corresponds to one Hamilton-Jacobi equation (i.5) for the Schrödinger equation. In this case the Schrödinger and Pauli equations describe, generally speaking, different quantum systems. The initial quantum system described by the Pauli equation has a potential energy $U$ and vector potential $\vec{\mathrm{A}}$. According to Theorem 2, a quantum system described by the Schrödinger equation has the potential energy

$$U_S = U + \delta U,\ \ \delta U \overset{\mathrm{def}}{=} \mathrm{P} + \delta\mathcal{V} - \delta\mathrm{Q} = \mathrm{P} + \hbar\kappa_1\kappa_2\left(\varphi_{21}\partial_t K + \frac{\hbar}{8m}\left|\nabla_r K\right|^2 + \frac{\hbar}{2m}\left|\nabla_r\varphi_{21}\right|^2\right), \tag{1.18}$$

and vector potential $\vec{\mathrm{A}}_{\Psi}$ (1.14). Nevertheless, both quantum systems have the same probability density function $f = |\psi_1|^2 + |\psi_2|^2$ and satisfy the same continuity equation, having the same probability current

$$\vec{J} = f\langle\vec{v}\rangle = f\left(\kappa_1\langle\vec{v}_1\rangle + \kappa_2\langle\vec{v}_2\rangle\right) = |\psi_1|^2\langle\vec{v}_1\rangle + |\psi_2|^2\langle\vec{v}_2\rangle. \tag{1.19}$$

Note that the quantum potential $\mathrm{Q}_S$ (1.17) in addition to the additive part $\mathrm{Q} = \kappa_1\mathrm{Q}_1 + \kappa_2\mathrm{Q}_2$ contains an additional correction $\delta\mathrm{Q}$, due to the mutual influence of the flows on each other. The situation is similar for the potential energy $U_S$ (1.18) and for the vector potential $\vec{\mathrm{A}}_{\Psi}$ (1.14), which contain additives $\delta U$ and $\vec{\mathrm{A}}_Q$ respectively.

## §2 Motion equations

Using the system of equations (1.4), we can construct for them corresponding system of equations.

**Theorem 3.** *Let the Hamilton-Jacobi equations (1.6) be fulfilled, then they correspond to the equations of motion for probability fluxes* $\langle\vec{v}_n\rangle$, $n = 1,2$

$$\frac{d_n}{dt}\langle\vec{v}_n\rangle \overset{\mathrm{def}}{=} \left(\partial_t + \langle\vec{v}_n\rangle\cdot\nabla_r\right)\langle\vec{v}_n\rangle = \frac{q}{m}\left(\vec{\mathrm{E}}_{\psi_n} + \langle\vec{v}_n\rangle\times\vec{\mathrm{B}}\right), \tag{2.1}$$

$$\vec{\mathrm{E}}_{\psi_n} \overset{\mathrm{def}}{=} -\partial_t\vec{\mathrm{A}} - \frac{1}{q}\nabla_r\mathcal{V}_n = \vec{\mathrm{E}} + \vec{\mathrm{E}}_{\mathrm{Q}_n} + \vec{\mathrm{E}}_{\mathrm{P}_n}, \tag{2.2}$$

$$\vec{\mathrm{E}} \overset{\mathrm{def}}{=} -\partial_t\vec{\mathrm{A}} - \frac{1}{q}\nabla_r U,\ \ \vec{\mathrm{E}}_{\mathrm{Q}_n} \overset{\mathrm{def}}{=} -\frac{1}{q}\nabla_r\mathrm{Q}_n,\ \ \vec{\mathrm{E}}_{\mathrm{P}_n} \overset{\mathrm{def}}{=} -\frac{1}{q}\nabla_r\mathrm{P}_n. \tag{2.3}$$

*The connection between electromagnetic fields (*$\vec{\mathrm{E}}_{\Psi}$, $\vec{\mathrm{B}}_{\Psi}$*) from the Schrödinger equation and fields (*$\vec{\mathrm{E}}$, $\vec{\mathrm{B}}$*) from the Pauli equation are defined by relation* $\vec{\mathrm{B}}_{\Psi} = \vec{\mathrm{B}} + \vec{\mathrm{B}}_Q$ *(1.9) and*

$$\vec{\mathrm{E}}_{\psi} \overset{\mathrm{def}}{=} \sum_{n=1}^{2} \kappa_n \vec{\mathrm{E}}_{\psi_n}, \qquad (2.4)$$

$$\vec{\mathrm{E}}_{\Psi} = \vec{\mathrm{E}}_{\psi} - \frac{m}{q}\left\{\frac{d\kappa_1}{dt} + \kappa_1\kappa_2\left[\left(\langle\vec{v}_2\rangle - \langle\vec{v}_1\rangle\right)\cdot\nabla_r\right]\right\}\left(\langle\vec{v}_2\rangle - \langle\vec{v}_1\rangle\right) - \langle\vec{v}\rangle\times\vec{\mathrm{B}}_Q, \qquad (2.5)$$

*where in accordance with (1.1)* $m\left(\langle\vec{v}_2\rangle - \langle\vec{v}_1\rangle\right) = \hbar\nabla_r\varphi_{21}$.

The proof of Theorem 3 is given in Appendix A.

Unlike the Schrödinger equation, there is no one equation of motion for the Pauli equation, but a system of two equations (2.1). This peculiarity is caused by the two-component (spinor) wave function $\psi$. The mathematical form of the motion equations (2.1) is similar to equation (i.7): the left side is the total time derivative of velocity, and the right side is proportional to force. Note that the total time derivatives $d_n/dt$ depend on the $n$-number index of the flux $\langle\vec{v}_n\rangle$. Let us recall, that for the Schrödinger equation there is only one flux $\langle\vec{v}\rangle$, hence only one total time derivative $d/dt$ (i.7) exists. By virtue of Lemma 1, the initial flux $\langle\vec{v}\rangle$ is divided into two fluxes $\langle\vec{v}_1\rangle$ and $\langle\vec{v}_2\rangle$. This gives us two total time derivatives $d_n/dt$ for each motion equation (2.1), corresponding to each flux.

The expressions for the force in the right-hand sides of equations (2.1) have an interesting form. On the one hand, this is the usual sum of the Coulomb force $q\vec{\mathrm{E}}_{\psi_n}$ and the Lorentz force $q\langle\vec{v}_n\rangle\times\vec{\mathrm{B}}$, acting on the $n$-th flux. On the other hand, the expression for the strength of the «electric» field $\vec{\mathrm{E}}_{\psi_n}$ has a significant difference from the strength $\vec{\mathrm{E}}_{\Psi}$. The field $\vec{\mathrm{E}}_{\psi_n}$ in addition to the electric field $\vec{\mathrm{E}}$ and field $\vec{\mathrm{E}}_Q$, due to the quantum pressure of the $n$-th flux, has a new additional term $\vec{\mathrm{E}}_{\mathrm{P}_n}$ (2.2)-(2.3). The field $\vec{\mathrm{E}}_{\mathrm{P}_n}$ is defined by the gradient of energy $\mathrm{P}_n = -\left(\vec{\mu}_{\pm}\cdot\vec{\mathrm{B}}\right)_n$, which is responsible for the interaction of the magnetic momentum $\vec{\mu}_{\pm}$ with field $\vec{\mathrm{B}}$ (see the Theorem 2). In a special case, where $|\psi_1| = |\psi_2|$ (metrics ${}_n g = \mathrm{I}_3$), force value is

$$\vec{F}_{\uparrow\downarrow} = q\vec{\mathrm{E}}_{\mathrm{P}_n} = \nabla_r\left({}_n g_{sk}\mu^s_{\uparrow\downarrow}\mathrm{B}^k\right)\Big|_{|\psi_1|=|\psi_2|} = \nabla_r\left(\vec{\mu}_{\uparrow\downarrow}\cdot\vec{\mathrm{B}}\right). \qquad (2.6)$$

Expression (2.6) is well known in physics as the force acting on a magnetic dipole $\vec{\mu}_{\uparrow\downarrow}$, placed in a magnetic field $\vec{\mathrm{B}}$. In general case, when $|\psi_1| \neq |\psi_2|$ the additional contribution to the force (2.6) will be made by the curvilinear metric ${}_n g \neq \mathrm{I}_3$. Recall that the covariant derivative of the metric is $\nabla_\ell\left({}_n g\right) = \left({}_n g\right)_{;\ell} = \partial_\ell\, {}_n g_{sk} - \Gamma^p_{\ell s}\, {}_n g_{pk} - \Gamma^p_{\ell k}\, {}_n g_{sp} = 0$, where $\Gamma^p_{sk}$ are the second order Christoffel symbols. According to (1.8) the radial components of the vector $\vec{\mu}_{\uparrow\downarrow} = -\mu_B\left(\cos\varphi_{21}\ \ \sin\varphi_{21}\ \ \pm 1\right)$ depend on the phase difference $\varphi_{21} = \varphi_2 - \varphi_1$ of wave functions $\psi_1, \psi_2$, which generally leads to the relatedness of the equations of motion (2.1) to each other, that is, to the impossibility of considering them separately. Thus, from a physical point of view, the new term $\vec{\mathrm{E}}_{\mathrm{P}_n}$ in the right side of equation (2.1) is responsible for the interaction of the particle's spin with external field.

Each flux $\langle \vec{v}_n \rangle$ has its own quantum pressure force $q\,\vec{E}_{Q_n} = \frac{\hbar^2}{2m}\nabla_r\left(\Delta_r|\psi_n|/|\psi_n|\right)$, since it is determined by its wave function amplitude $|\psi_n|$. As for the Schrödinger equation the force of quantum pressure $q\,\vec{E}_{Q_n}$ counteracts the external force $-\nabla_r U$. The total electric force $q\,\vec{E}_{\psi_n}$ in the right-hand side of the equation (2.1) is determined by the field $\vec{E}_{\psi_n}$ (2.2), which is different from the external electric field $\vec{E}$. According to (2.5) the superposition (2.4) of fields $\vec{E}_{\psi_n}$ with expansion coefficients $\kappa_n$ is expressed in terms of the field $\vec{E}_\Psi$ for the Schrödinger equation. The value $\vec{B}_Q$ in explicit form is known from the expression (1.3). Note that in the absence of a coordinate dependence, the phase difference $\varphi_{21} = \varphi_{21}(t)$ in expression (2.5) is significantly simplified into $\vec{E}_\Psi = \vec{E}_\psi$. In this case, the derivative of $\vec{\mu}_{\uparrow\downarrow}$ with respect to the coordinate $\vec{r}$ in the expression for the force (2.6) is absent, that is

$$\vec{F}_{\uparrow\downarrow} = \left(\vec{\mu}_{\uparrow\downarrow}\cdot\nabla_r\right)\vec{B} + \vec{\mu}_{\uparrow\downarrow}\times \operatorname{curl}_r \vec{B}. \tag{2.7}$$

Theorem 3 establishes the motion equations of the probability fluxes $\langle \vec{v}_n \rangle$ in electromagnetic fields $\vec{E}$, $\vec{B}$. From a physical point of view the fields $\vec{E}$, $\vec{B}$ must satisfy Maxwell's equations. According to Theorem 3, the quantum system described by the Pauli equation is associated with the quantum system described by the Schrödinger equation with the fields $\vec{E}_\Psi$, $\vec{B}_\Psi$. A natural question arises: «Under what conditions do fields $\vec{E}_\Psi$, $\vec{B}_\Psi$ satisfy Maxwell's equations?»

**Definition 3.** *Let us define the electromagnetic fields $\vec{D}_I, \vec{H}_I$ and corresponding charge density $\rho_I$*

$$\vec{D}_I \overset{\text{def}}{=} \varepsilon_0\,\vec{E}_I,\ \vec{B}_I \overset{\text{def}}{=} \mu_0\,\vec{H}_I,\ \rho_I \overset{\text{def}}{=} \operatorname{div}_r \vec{D}_I,\ I = \{S, Q, \Psi, \varnothing\}, \tag{2.8}$$

*where $(\varepsilon_0\mu_0)^{-2} = c$ is the speed of light.*

**Theorem 4.** *Let the Maxwell equations in the Lorenz gauge be satisfied for the electromagnetic fields $\vec{E}, \vec{B}$ of the quantum system, described by the Pauli equation*

$$\Box\mathcal{A}^\nu = \mu_0\mathcal{J}^\nu,\ \partial_\nu\mathcal{A}^\nu = 0,\ \mathcal{A}^\nu = \left(\varphi/c,\ \vec{A}\right),\ \mathcal{J}^\nu = \left(c\rho,\ \vec{J}\right), \tag{2.9}$$

*and for the 4-potential $\mathcal{A}^\nu_\Psi = \left(\varphi_\Psi/c,\ \vec{A}_\Psi\right)$, corresponding to the Schrödinger equation, and let the Lorenz $\Psi$-gauge $\partial_\nu\mathcal{A}^\nu_\Psi = 0$ be fulfilled and*

$$\Box\vec{A}_Q = \mu_0\,\vec{J}_Q, \tag{2.10}$$

*then $\vec{E}_\Psi, \vec{B}_\Psi$ satisfy the following Maxwell equations*

$$\Box \mathcal{A}_{\Psi}^{\nu} = \mu_0 \mathcal{J}_{\Psi}^{\nu}, \qquad \mathcal{J}_{\Psi}^{\nu} = \left(c\rho_{\Psi},\, \vec{\mathrm{J}}_{\Psi}\right), \qquad \rho_{\Psi} = \varepsilon_0 \Box \varphi_{\Psi}, \quad \vec{\mathrm{J}}_{\Psi} = \vec{\mathrm{J}} + \vec{\mathrm{J}}_Q, \tag{2.11}$$

*in this case the following relations are valid*

$$q\varphi_{\Psi} = \mathrm{V}_{\Psi}, \qquad q\varphi = U, \qquad q\,\varphi_Q = \mathrm{Q}_S, \qquad q\delta\varphi = \delta U, \tag{2.12}$$

$$\varphi_{\Psi} = \varphi_S + \varphi_Q, \qquad \varphi_S = \varphi + \delta\varphi, \tag{2.13}$$

$$\rho_{\Psi} = \rho_S + \rho_Q, \qquad \rho_S = \rho + \delta\rho, \tag{2.14}$$

$$\rho_{\Psi} = \varepsilon_0 \Box \varphi_{\Psi}, \quad \rho = \varepsilon_0 \Box \varphi, \qquad \delta\rho = -\varepsilon_0 \Delta_r \delta\varphi, \tag{2.15}$$

$$\rho_S = \varepsilon_0 \Box \varphi_S - \varepsilon_0 \partial_0 \partial^0 \delta\varphi, \qquad \rho_Q = \varepsilon_0 \Box \varphi_Q + \varepsilon_0 \partial_0 \partial^0 \delta\varphi, \tag{2.16}$$

$$\vec{\mathrm{E}}_{\Psi} = \vec{\mathrm{E}}_S + \vec{\mathrm{E}}_Q = \vec{\mathrm{E}} - \partial_t \vec{\mathrm{A}}_Q - \nabla_r\left(\varphi_Q + \delta\varphi\right), \qquad \vec{\mathrm{E}}_S = \vec{\mathrm{E}} - \nabla_r \delta\varphi, \qquad \vec{\mathrm{B}}_{\Psi} = \vec{\mathrm{B}} + \vec{\mathrm{B}}_Q, \tag{2.17}$$

*where* $\Box = \partial_\mu \partial^\mu = g_{\nu\mu} \partial^\nu \partial^\mu$ *is the d'Alembert operator and* $g_{\nu\mu} \mapsto (+,-,-,-)$ *is a metric of space-time.*

Proof of the theorem 4 is given in Appendix A.

**Remark 2.** Let us notice that field $\vec{\mathrm{A}}_Q$ is known in an explicit form (Lemma 1 (1.3)), thus the current density $\vec{\mathrm{J}}_Q$ (2.10) can be found. Therefore, using $\vec{\mathrm{J}}$ one can find $\vec{\mathrm{J}}_{\Psi}$ (2.11). The continuity equations $\partial_\nu \mathcal{J}^{\nu} = 0$ and $\partial_\nu \mathcal{J}_{\Psi}^{\nu} = 0$ containing currents $\vec{\mathrm{J}}$ and $\vec{\mathrm{J}}_{\Psi}$ follow from the Maxwell equations for fields $\vec{\mathrm{D}}$ and $\vec{\mathrm{D}}_{\Psi}$. In the case of self-consistent system when $q|\Psi|^2 = \rho_{\Psi}$ the current densities $\vec{\mathrm{J}}_{\Psi}$ and $\vec{J}$ can be equal, that is $\vec{\mathrm{J}}_{\Psi} = \vec{J} = f\langle\vec{v}\rangle$. Examples of such systems are considered in [15, 36].

There is a special case when the Lorenz $\Psi$-gauge (i.12) for the Schrödinger equation is a sum of two gauges (i.14). But in the general case this symmetry according to Theorem 4 is violated (A.53), since

$$c^2 \operatorname{div}_r \vec{\mathrm{A}}_Q = -\partial_t \varphi_Q - \partial_t \delta\varphi \neq -\partial_t \varphi_Q. \tag{2.18}$$

As it can be seen from (2.18), the underlying problem is in term $\delta\varphi$. If the $\delta\varphi$ is independent of time $t$, then symmetry is restored. For example, according to (1.18) $\partial_t \delta\varphi = 0$ for time independent potentials $\varphi$ and $\varphi_S$. In this case using (2.16) the equations for potentials $\varphi_S$ and $\varphi_Q$ take the simplest form $\rho_S = \varepsilon_0 \Box \varphi_S$ and $\rho_Q = \varepsilon_0 \Box \varphi_Q$.

Note that the motion equations (2.1) are in strict agreement with the motion equation (i.7). Indeed, if we substitute the velocity decomposition $\langle\vec{v}\rangle$ via velocities $\langle\vec{v}_n\rangle$ (1.1) and expressions for fields $\vec{\mathrm{E}}_{\Psi}$, $\vec{\mathrm{B}}_{\Psi}$ via fields $\vec{\mathrm{E}}_{\psi_n}$ (2.2) and $\vec{\mathrm{B}}$ (2.17) into equation (i.7), then we obtain expression (i.7) to be fulfilled.

Each group of fields $\vec{\mathrm{E}}_{\Psi}$, $\vec{\mathrm{B}}_{\Psi}$ (potentials $U_S, \vec{\mathrm{A}}_{\Psi}$) or $\vec{\mathrm{E}}$, $\vec{\mathrm{B}}$ (potentials $U, \vec{\mathrm{A}}$) corresponds to its own physical system described by equations (i.3), (i.5), (i.7) or (i.17), (1.6), (2.1), which are in a strict mathematical relationship as described above.

## §3 Exact solutions of the Pauli equation

We demonstrate the results of the theorems from §1-2 on the exact solution of the Pauli equation. As an example, let us consider a quantum system with a uniform magnetic field and an asymmetric quadratic potential. We start with the consequence of Lemma 1 and Theorems 1-4 for the case of a uniform magnetic field.

**Corollary 1.** *Let the vector potential be defined by constant values* $b_2 > b_1$ *and takes the following form* $\vec{\mathrm{A}}(x,y)=(b_1 y\ b_2 x\ 0)$, *and* $q=-e$, *then* $\kappa_n = |c_n|^2$, $|\Psi|^2 = |\psi|^2 = |\Upsilon|^2$ *and* $\mathrm{B} = b_2 - b_1$, *where*

$$\psi(\vec{r},t)=\Upsilon(\vec{r},t)\chi(t),\ \chi(t)=\begin{bmatrix} c_1 e^{-i\omega_0 t} \\ c_2 e^{i\omega_0 t}\end{bmatrix},\ |c_1|^2+|c_2|^2=1,\ \omega_0=\frac{e\mathrm{B}}{2m},\ \mu_B\mathrm{B}=\hbar\omega_0, \quad (3.1)$$

*and the wave function* $\Upsilon$ *is a solution of the Schrödinger equation*

$$i\hbar\partial_t\Upsilon=\frac{1}{2m}\left(\hat{\mathrm{p}}-q\vec{\mathrm{A}}\right)^2\Upsilon+U\Upsilon. \quad (3.2)$$

*In this case, the expressions for potentials and fields take the form*

$$\mathrm{P}_n=-(-1)^n\mu_B\mathrm{B}, \qquad \mathrm{P}=(\kappa_1-\kappa_2)\mu_B\mathrm{B}, \quad (3.3)$$

$$\delta\mathrm{Q}=0,\ \mathrm{Q}_S=\mathrm{Q}=\mathrm{Q}_n, \qquad \delta\mathcal{V}=0,\ \mathrm{V}_\Psi=\mathcal{V}, \qquad \delta U=\mathrm{P},\ U_S=U+(\kappa_1-\kappa_2)\mu_B\mathrm{B}, \quad (3.4)$$

$$\mathcal{V}_n=\mathrm{V}_\Psi+\left[\kappa_2-\kappa_1-(-1)^n\right]\mu_B\mathrm{B}, \quad (3.5)$$

$$\vec{\mathrm{A}}_Q=\vec{\theta},\ \vec{\mathrm{A}}_\Psi=\vec{\mathrm{A}}, \qquad \vec{\mathrm{B}}_Q=\vec{\theta},\ \vec{\mathrm{B}}_\Psi=\vec{\mathrm{B}}, \quad (3.6)$$

$$\vec{\mathrm{E}}_{\psi_n}=\vec{\mathrm{E}}+\vec{\mathrm{E}}_{\mathrm{Q}}=\vec{\mathrm{E}}_\Psi=\vec{\mathrm{E}}_\psi, \qquad \vec{\mathrm{E}}=\vec{\mathrm{E}}_S,\ \vec{\mathrm{E}}_{\mathrm{Q}_n}=\vec{\mathrm{E}}_{\mathrm{Q}},\ \vec{\mathrm{E}}_{\mathrm{P}_n}=0. \quad (3.7)$$

*For phases of wave functions* $\psi_n$ *and* $\Psi$ *the following relations are valid*

$$\varphi_n=\varphi_\Upsilon+\varphi_{c_n}+(-1)^n\omega_0 t, \qquad \varphi=\varphi_\Upsilon+\omega_0 t(\kappa_2-\kappa_1)+\kappa_1\varphi_{c_1}+\kappa_2\varphi_{c_2}+\pi l, \quad (3.8)$$

*where* $\bar{\mathrm{S}}_z=\chi^\dagger\hat{\mathrm{S}}_z\chi=\frac{\hbar}{2}(\kappa_1-\kappa_2)$, $\hat{\mathrm{S}}=\frac{\hbar}{2}\hat{\vec{\sigma}}$ *is a spin operator, and* $\varphi_\Upsilon$ *is a phase of the wave function* $\Upsilon$ *and* $\varphi_{c_n}$ *are phases of* $c_n$. *The function of probability density* $f_n=|\psi_n|^2$ *and* $f=|\Psi|^2=|\Upsilon|^2$ *satisfy the continuity equations*

$$\partial_t f+\mathrm{div}_r\left(f\langle\vec{v}\rangle\right)=0, \qquad \partial_t f_n+\mathrm{div}_r\left(f_n\langle\vec{v}_n\rangle\right)=0. \quad (3.9)$$

*If* $\varphi_\Upsilon=-\frac{\mathcal{E}}{\hbar}t$, *then* $\langle\vec{v}_n\rangle=\frac{e}{m}\vec{\mathrm{A}}=\langle\vec{v}\rangle$. *The Hamilton-Jacobi equations (i.5) and (1.6) take the form*

$$\mathcal{E}-\hbar\omega_0(\kappa_2-\kappa_1)=\frac{1}{2m}\left|e\vec{\mathrm{A}}\right|^2+\mathrm{V}_\Psi, \quad (3.10)$$

$$\mathcal{E}-(-1)^n\hbar\omega_0=\frac{1}{2m}\left|e\vec{\mathrm{A}}\right|^2+\mathrm{V}_\Psi+\left[\kappa_2-\kappa_1-(-1)^n\right]\mu_B\,\mathrm{B}, \tag{3.11}$$

*and the corresponding motion equations (2.1) become correct identities.*

Proof of Corollary1 is given in Appendix B.

**Remark 3.** Note that the sum of equations (3.11) gives equation (3.10). A similar result is obtained if we consider that $\mu_B\,\mathrm{B}=\hbar\omega_0$. In this case, equation (3.11) becomes equation (3.10). The right-hand side of equation (3.10) is the Hamiltonian function $H$, which is the sum of kinetic and potential energy. Thus, the total energy $H=\mathcal{E}+2\overline{\hat{\mathrm{S}}_z}\,\omega_0$. The value $\overline{\hat{\mathrm{S}}_z}$ is determined by the expansion coefficients $\kappa_n$ of the probability flux velocity $\langle\vec{v}\rangle=\kappa_1\langle\vec{v}_1\rangle+\kappa_2\langle\vec{v}_2\rangle$ (1.1). The coefficient $\kappa_n$ defines the probability fraction for $n$-th flux. According to (1.2), if $\kappa_1=1\Rightarrow\kappa_2=0$, that is $\overline{\hat{\mathrm{S}}_z}=+1/2$ and $H_+=\mathcal{E}+\hbar\omega_0$. If $\kappa_1=0\Rightarrow\kappa_2=1$ and $\overline{\hat{\mathrm{S}}_z}=-1/2$, then $H_-=\mathcal{E}-\hbar\omega_0$. The given expressions for energy $H$ correspond to the Zeeman effect on the splitting of energy levels. Note that the Hamilton-Jacobi equation (3.10) can be rewritten in a simple form according to (3.4) $\mathcal{E}=\frac{1}{2m}\left|e\vec{\mathrm{A}}\right|^2+U_S+\mathrm{Q}_S$.

**Theorem 5.** *Let the vector potential* $\vec{\mathrm{A}}$ *has only an angular axial component* $\vec{\mathrm{A}}=\tfrac{1}{2}\mathrm{B}\,\rho\vec{e}_\phi$, *that is* $b_2=-b_1=\mathrm{B}/2$, *then the spinor* $\psi_{s,\ell,l}$

$$\psi_{s,\ell,l}(\rho,\phi,z,t)=N_{s,\ell,l}\rho^{|\ell|}e^{-\frac{1}{2}\xi^2(\rho^2+z^2)+i\ell\phi-i\frac{\mathcal{E}_{s,\ell,l}}{\hbar}t}L_s^{(|\ell|)}(\xi^2\rho^2)H_l(\xi z)\chi(t), \tag{3.12}$$

$$\mathcal{E}_{s,\ell,l}=\hbar\omega_0\ell+\hbar\Omega\left(|\ell|+l+2s+\frac{3}{2}\right), \tag{3.13}$$

$$\Omega^2=\omega^2+\omega_0^2,\ \xi^2=\frac{m\Omega}{\hbar},\ N_{s,\ell,l}^2=\frac{s!\,\xi^{2|\ell|+3}}{2^l l!\,\pi^{3/2}\Gamma(s+|\ell|+1)},\ \ell\in\mathbb{Z}, \tag{3.14}$$

*is a solution of the Pauli equation with potential* $U$ *and quantum potential* $\mathrm{Q}$

$$U(\rho,z)=\frac{m\omega^2\rho^2}{2}+\frac{m\Omega^2z^2}{2}, \tag{3.15}$$

$$\mathrm{Q}_S=\mathrm{Q}_\Upsilon=\mathrm{Q}_n=\mathrm{Q}_{s,\ell,l}=\hbar\Omega\left(|\ell|+l+2s+\frac{3}{2}\right)-\frac{m\Omega^2r^2}{2}-\frac{\hbar^2\ell^2}{2m\rho^2}, \tag{3.16}$$

*where* $L_s^{(|\ell|)}$ *are the generalized Laguerre polynomials,* $H_l$ *are the Hermite polynomials,* $r^2=\rho^2+z^2$. *Vector fields of probability flux have the form*

$$\langle\vec{v}\rangle_\ell(\rho)=\langle\vec{v}_n\rangle_\ell(\rho)=\left(\frac{\hbar\ell}{m\rho}+\omega_0\rho\right)\vec{e}_\phi, \tag{3.17}$$

*and satisfy the Hamilton-Jacobi equation*

$$\mathcal{E}_{s,\ell,l}=\frac{m}{2}\left|\langle\vec{v}\rangle_\ell\right|^2+U+\mathrm{Q}_{s,\ell,l}. \tag{3.18}$$

The proof of Theorem 5 is given in Appendix B.

Indices $(s,\ell,l)$ designate a certain quantum state with energy (3.13). The phase $\varphi_\Upsilon$, and therefore phases $\varphi,\varphi_n$ (3.8) depend not only on time, but also on the polar angle $\phi$. Such dependence is a special case of the so-called $\Psi$ - model of micro and macro systems [13], in which

$$\varphi_{\Psi-\text{model}}(\theta,\phi,t)=\ell\phi+j\theta-\frac{\mathcal{E}}{\hbar}t+2\pi k,\ \ell,j,k\in\mathbb{Z}, \tag{3.19}$$

where $(\theta,\phi)$ are the angles in the spherical coordinate system. If $j=0$ the phase (3.19) goes to

$$\varphi_\Upsilon(\phi,t)=\varphi_{\Psi-\text{model}}(\theta,\phi,t)\Big|_{j=0}=\ell\phi-\frac{\mathcal{E}}{\hbar}t+2\pi k. \tag{3.20}$$

Index $\ell$ determines the component $\langle\vec{v}_\phi\rangle_\ell=\frac{\hbar\ell}{m\rho}\vec{e}_\phi$ of the vortex field (3.17). Recall that the value $\hbar\ell$ corresponds to the axial angular momentum of the system $\vec{\mathrm{M}}_\ell=\vec{\rho}\times\langle\vec{p}_\phi\rangle_\ell=\vec{e}_z\hbar\ell$, where $\vec{\rho}=\vec{e}_\rho\rho$ and $\langle\vec{p}_\phi\rangle_\ell=m\langle\vec{v}_\phi\rangle_\ell$ is a momentum. Despite the dependence of phase (3.20) on the angle $\phi$ the time derivative in the Hamilton-Jacobi equation remains unchanged $-\hbar\partial_t\varphi_\Upsilon=\mathcal{E}$. An important remark should be made here. Theorem 5 considers three types of the Hamilton-Jacobi equations corresponding to wave functions $\Psi$, $\psi$ and $\Upsilon$:

$$-\hbar\partial_t\varphi=\mathcal{E}_{s,\ell,l}+2\bar{\hat{\mathrm{S}}}_z=\frac{m}{2}\left|\langle\vec{v}\rangle_\ell\right|^2+\mathrm{V}_\Psi,\ \mathrm{V}_\Psi=U_S+\mathrm{Q}_S=U+2\bar{\hat{\mathrm{S}}}_z+\mathrm{Q}_{s,\ell,l}, \tag{3.21}$$

$$-\hbar\partial_t\varphi_\Upsilon=\mathcal{E}_{s,\ell,l}=\frac{m}{2}\left|\langle\vec{v}\rangle_\ell\right|^2+\mathrm{V}_\Upsilon,\ \mathrm{V}_\Upsilon=U+\mathrm{Q}_\Upsilon=U+\mathrm{Q}_{s,\ell,l}, \tag{3.22}$$

$$-\hbar\partial_t\varphi_n=\mathcal{E}_{s,\ell,l}-(-1)^n\hbar\omega_0=\frac{m}{2}\left|\langle\vec{v}\rangle_\ell\right|^2+\mathcal{V}_n,\ \mathcal{V}_n=U+\mathrm{Q}_{s,\ell,l}-(-1)^n\mu_B\mathrm{B}, \tag{3.23}$$

where expressions (i.5), (3.5) and (3.8) are taken into account. All three equations (3.21)-(3.23) are completely identical and equivalent to equation (3.18). However, the energy level splitting (Remark 3) is only observed in equations (3.21) and (3.23), while it is naturally absent in equation (3.22).

Note that the formal theoretical limit (the speed of light $c$) of the vortex component $\frac{\hbar|\ell|}{m\rho_\ell}=c$ is reached on the distances $\rho_\ell=\lambda\!\!\!^{-}_C|\ell|$, where $\lambda\!\!\!^{-}_C$ is the reduced Compton wavelength.

The motion equation (i.7) for the flux with velocity $\langle\vec{v}\rangle_\ell$ reduces to the well-known expression for the centripetal force

$$\vec{F}_\ell=m\frac{d}{dt}\langle\vec{v}\rangle_\ell=m\left(\langle\vec{v}\rangle_\ell\cdot\nabla_r\right)\langle\vec{v}\rangle_\ell=-\vec{e}_\rho\frac{m}{\rho}\left|\langle\vec{v}\rangle_\ell\right|^2=-\vec{e}_\rho\left(\frac{\hbar^2\ell^2}{m\rho^3}+2\ell\frac{\hbar\omega_0}{\rho}+m\omega_0^2\rho\right), \tag{3.24}$$

where under $\ell = 0$ there is only a linear angular velocity $\langle \vec{v} \rangle_0 = \vec{e}_\phi \omega_0 \rho$ and the force $\vec{F}_0 = -\vec{e}_\rho m \omega_0^2 \rho$, which is well-known from the general physics course.

Depending on the sign of $\ell$ the flux velocity (3.17) will strengthen or weaken. Fig. 1 shows the distribution (3.17) for three cases $\ell = \pm 1$ and $\ell = 0$. White arrows in fig. 1 illustrate the directions of the fluxes. Red color in the fig. 1 corresponds to the maximum, and blue one – to the minimum of velocity absolute value $\left| \langle \vec{v} \rangle_\ell \right|$. The axes in fig. 1 stand for the coordinates $(x, y)$ in a plane, which is transverse to the magnetic field direction. The directions of the fluxes' velocities $\langle \vec{v} \rangle_\ell$ are shown with multi-colored dashes in fig. 1. The length of the multi-colored dashes is proportional to the velocity absolute value $\left| \langle \vec{v} \rangle_\ell \right|$ and actually corresponds to the graph color.

The component $\frac{e}{m} \vec{\mathrm{A}} = \omega_0 \rho \vec{e}_\phi$ grows linearly with the radius $\rho$, therefore, its maximum value is reached at the outer radii. Component $\langle \vec{v}_\phi \rangle_\ell$ reaches its maximum in the central region and decreases with increasing radius $\rho$. In fig. 1 on the left ($\ell = -1$) the fluxes are directed towards each other, so a radius $\rho_0 = \sqrt{-\hbar \ell / m \omega_0}$ exists, for which $\langle \vec{v} \rangle_\ell (\rho_0) = \vec{\theta}$. A circle $\rho = \rho_0$ represented with a white dashed line on the left of fig. 1. The resulting force $\vec{F}_\ell = \vec{\theta}$ and the Lorentz force $e \langle \vec{v} \rangle_\ell \times \vec{\mathrm{B}} = \vec{\theta}$ on the circle $\rho = \rho_0$, according to (3.24). Consequently, there is a balance between the external force $-\nabla_r U$ and the force of quantum pressure $-\nabla_r \mathrm{Q}$, that is $\nabla_r (U + \mathrm{Q}) = \vec{\theta}$.

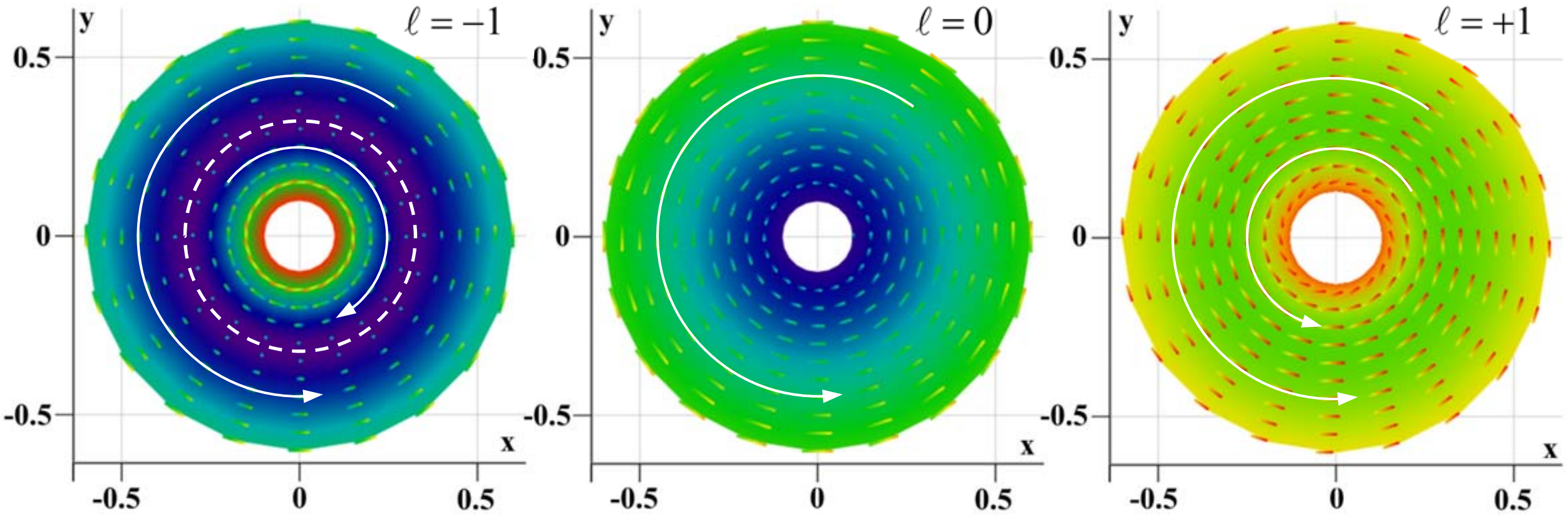

Fig. 1 Probability flux velocity distribution $\langle \vec{v} \rangle_\ell$.

In the center graph of fig. 1 $\ell = 0$, therefore, there is only an external flow that grows linearly with the radius $\frac{e}{m} \vec{\mathrm{A}}$. Since $\ell = +1$ on the right graph of fig. 1, then the external and internal flows add up, resulting in an increased total flux $\langle \vec{v} \rangle_\ell$ (compare the colors of the left and right graphs in fig. 1).

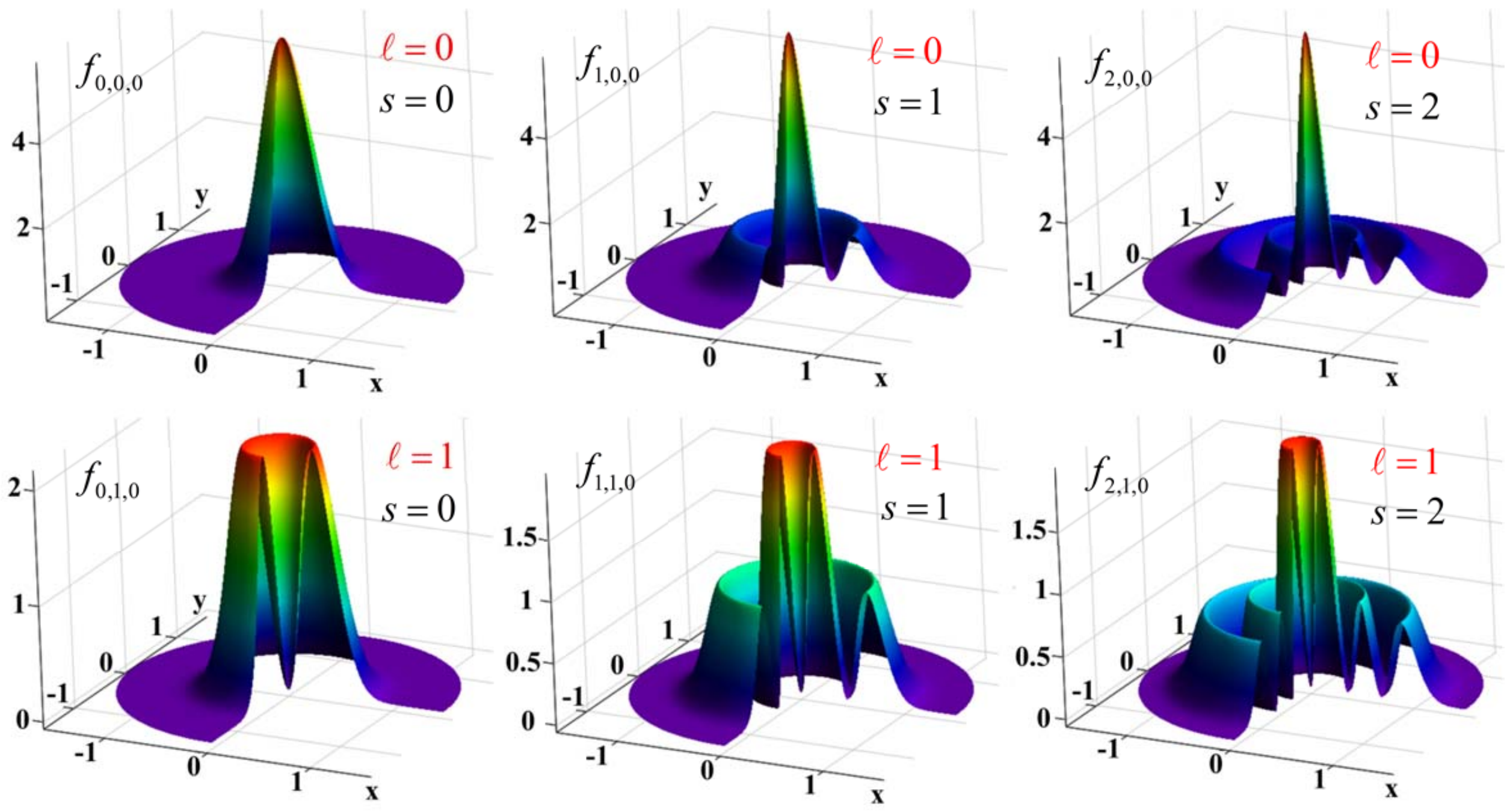

Fig. 2 Probability density distribution $f_{s,\ell,0}$ by $z = 0$.

Fig. 2 shows the probability density distributions $f_{s,\ell,0} = |\psi_{s,\ell,0}|^2$ by $z = 0$ in the transverse plane. Distributions $f_{s,\ell,l}$ are azimuthally symmetric, so for better clarity, fig. 2 shows these distributions for angels $0 \le \phi \le 3\pi/2$. We use the following values $\ell = \{0,1\}$ and $s = \{0,1,2\}$. The top three graphs in fig. 2 correspond to $\ell = 0$, and the bottom three ones to $\ell = 1$. The functions $f_{s,\ell,0}$ are plotted in accordance with (3.12). Each density $f_{s,\ell,0}$ in fig. 2 corresponds to its energy value $\mathcal{E}_{s,\ell,0}$ (3.13).

Note that according to (3.12), the distributions $f_{s,\ell,l}$ do not depend on the sign $\ell$, therefore, the bottom three graphs in fig. 2 are also valid for $f_{s,-1,0}$. The velocity field $\langle \vec{v} \rangle_0$ in fig. 1 (the center) correspond the top three graphs to $f_{s,0,0}$ in fig. 2. And the velocity distributions $\langle \vec{v} \rangle_{\pm 1}$ (see the left and right graphs in fig. 1) will correspond the three lower graphs of density $f_{s,1,0}$ in fig. 2. Let us note that the velocity field (3.17) does not depend on the state number $s$ and depends significantly on the magnitude and the sign of the azimuthal number $\ell$. The probability density $f_{s,\ell,l}$ does not depend on the sign of the azimuthal number $\ell$, but it depends on the values $s, \ell, l$.

If the distributions $f_{s,\ell,l}$ and $\langle \vec{v} \rangle_{s,\ell}$ are known then the magnetic moment of a quantum system can be found from Theorem 5

$$\vec{M}_{s,\ell,l} = \frac{1}{2} \int_{\mathbb{R}^3} \left( \vec{r} \times \vec{J}_{s,\ell,l} \right) d^3 r = -\frac{e}{2} \int_{\mathbb{R}^3} |\psi_{s,\ell,l}|^2 \, \vec{r} \times \langle \vec{v} \rangle_\ell \, d^3 r. \tag{3.25}$$

The direct calculation of the integral (3.25) leads to the following result (Appendix B)

$$\vec{M}_{s,\ell} = -\mu_B \left[ \ell + \frac{\omega_0}{\Omega}\left(|\ell| + 2s + 1\right) \right] \vec{e}_z. \qquad (3.26)$$

Note that the magnetic moment (3.26) has only an axial component along the external magnetic field $\vec{\mathrm{B}}$ and consists of two terms. The first term $\mu_B \ell$ in (3.26) is related to the vortex component of the probability flux velocity $\langle \vec{v}_\phi \rangle_\ell$ and the second term is related to the external field $\frac{e}{m}\vec{\mathrm{A}}$. When $\ell > 0$ the magnetic momentum $\vec{M}_{s,\ell}$ is in the opposite direction to the external magnetic field $\vec{\mathrm{B}}$ ($\vec{\mathrm{B}} \uparrow\downarrow \vec{M}_{s,\ell}$). If the value $\ell < 0$ then it is possible to define the frequency $\omega$ of the external electric potential $U$ (3.15) in such a way that $\vec{M}_{s,\ell}$ will be directed along the field $\vec{\mathrm{B}}$ ($\vec{\mathrm{B}} \uparrow\uparrow \vec{M}_{s,\ell}$). Indeed, when $\omega > 0$

$$\ell_s = -\frac{2s+1}{1 - \omega_0/\Omega} \Rightarrow \vec{M}_{s,\ell_s} = \vec{\theta}, \qquad (3.27)$$

where $\omega_0/\Omega < 1$ according to (3.14). If $\omega = 0$ then $\vec{M}_{s,\ell_s} \neq \vec{\theta}$. Since $\ell \in \mathbb{Z}$, there must be a certain ratio between the frequencies $\omega_0$ and $\omega$. The frequency $\omega_0$ is determined by the external magnetic field $\vec{\mathrm{B}}$, and $\omega$ is connected to the electric potential $U$. For example, in the simplest case, we obtain

$$\omega = \omega_0 \frac{\sqrt{2j-1}}{j-1} \Rightarrow \ell_s = -j(2s+1), \; j = 2,3,... \qquad (3.28)$$

Thus, if condition (3.28) is satisfied then $\vec{M}_{s,\ell_s} = \vec{\theta}$ for any number of the quantum state $s$. Note that condition (3.28) is somewhat similar to the condition for magnetic resonance, where the selection of the frequency of an external magnetic field leads to the spin reversal.

**Conclusion**

In historical retrospect the Pauli equation was the first fundamental equation that took spin into account when describing a quantum system. Numerous attempts to understand the nature of spin from the perspective of classical mechanics have encountered significant difficulties. These facts have sparked the interest of the authors in studying the kinematic properties of the Pauli equation.

The Pauli and Schrödinger equations are non-relativistic equations, which makes it possible to study their properties in terms of non-relativistic classical continuum mechanics. It should be noted that from the standpoint of the Vlasov chain of equations the function $f(\vec{r},t)$ can be interpreted in different ways depending upon a certain problem. For example, in continuum mechanics $f(\vec{r},t)$ it can correspond to the density of matter or charge, and in quantum mechanics, it relates to the density of probabilities for a single particle. From a mathematical perspective, the fundamental equations of motion have a similar form. This fact allows us to match exact solutions for one physical system with exact solutions for another physical system.

It is important to note that, in general, this paper does not consider the Schrödinger equation as a partial case of the Pauli equation. Instead, we connect the Schrödinger equation with the Pauli equation through the first Vlasov equation. In the example discussed in §3, the way of solution of the Pauli equation reduces to solving the Schrödinger equation for the wave function $\Upsilon$. But the function $\Upsilon$, generally speaking, differs from the function $\Psi$ by its phase. The phase of the wave function is important in this example, as it sets different energies in the Hamilton-Jacobi equation according to (3.8) and (3.20), resulting in splitting of the energy levels. In general, when the phase depends on the coordinate, this difference can lead to different average velocities $\langle \vec{v} \rangle$ or $\langle \vec{v}_n \rangle$ (1.1). Therefore, in this consideration, the Schrödinger and Pauli equations describe different quantum systems (with different electromagnetic fields), but they are connected through the first Vlasov equation.

Note that knowing the wave functions of the Schrödinger and Pauli equations allows us to construct the Wigner functions and proceed to the second Vlasov equation.

**Appendix A**

***Proof of Lemma 1.***

Let us transform the expression (i.16)

$$\psi^\dagger \mathrm{I}\nabla_r\psi - \psi^T \mathrm{I}\nabla_r\psi^* = \psi_1^*\nabla_r\psi_1 + \psi_2^*\nabla_r\psi_2 - \psi_1\nabla_r\psi_1^* - \psi_2\nabla_r\psi_2^* =$$

$$= \psi_1\psi_1^*\left(\frac{\nabla_r\psi_1}{\psi_1} - \frac{\nabla_r\psi_1^*}{\psi_1^*}\right) + \psi_2\psi_2^*\left(\frac{\nabla_r\psi_2}{\psi_2} - \frac{\nabla_r\psi_2^*}{\psi_2^*}\right) = |\psi_1|^2\nabla_r \mathrm{Ln}\left(\frac{\psi_1}{\psi_1^*}\right) + |\psi_2|^2\nabla_r \mathrm{Ln}\left(\frac{\psi_2}{\psi_2^*}\right),$$

$$\psi^\dagger \mathrm{I}\nabla_r\psi - \psi^T \mathrm{I}\nabla_r\psi^* = i\left(|\psi_1|^2\nabla_r\Phi_1 + |\psi_2|^2\nabla_r\Phi_2\right), \tag{A.1}$$

where expression (1.2) is taken into account. Substituting (A.1) into (i.16), we obtain

$$\langle \vec{v} \rangle = -\alpha\left(\kappa_1\nabla_r\Phi_1 + \kappa_2\nabla_r\Phi_2\right) + \gamma\vec{\mathrm{A}} = \kappa_1\left(-\alpha\nabla_r\Phi_1 + \gamma\vec{\mathrm{A}}\right) + \kappa_2\left(-\alpha\nabla_r\Phi_2 + \gamma\vec{\mathrm{A}}\right), \tag{A.2}$$

where the relations (1.2) are taken into account for $\kappa_n$. Let us reduce the decomposition (A.2) to the form of the Helmholtz decomposition into a potential and a vortex field. Note that according to the definition (1.2), the following expression is valid

$$\sum_{n=1}^{2}\kappa_n\nabla_r\Phi_n = \nabla_r\Phi - \sum_{n=1}^{2}\Phi_n\nabla_r\kappa_n. \tag{A.3}$$

Substituting (A.3) into (A.2) we obtain the Helmholtz decomposition

$$\langle \vec{v} \rangle = -\alpha\nabla_r\Phi + \alpha\sum_{n=1}^{2}\Phi_n\nabla_r\kappa_n + \gamma\vec{\mathrm{A}} \Rightarrow \vec{\mathrm{A}}_\Psi = \vec{\mathrm{A}} + \frac{\alpha}{\gamma}\sum_{n=1}^{2}\Phi_n\nabla_r\kappa_n. \tag{A.4}$$

From expressions (A.2) and (A.4), the validity of representation (1.1) and expressions (1.3) follows. The lemma is proved.

***Proof of Theorem 1.***

We can transform the Pauli equation into a system of two equations using the relation $(\vec{\sigma}\cdot\vec{a})(\vec{\sigma}\cdot\vec{b}) = \mathrm{I}(\vec{a}\cdot\vec{b}) + i\vec{\sigma}\cdot(\vec{a}\times\vec{b})$

$$\frac{i}{\beta}\begin{pmatrix}\partial_t\psi_1\\ \partial_t\psi_2\end{pmatrix} = -\alpha\beta\mathrm{I}\left(\hat{\mathrm{p}} - \frac{\gamma}{2\alpha\beta}\vec{\mathrm{A}}\right)^2\begin{pmatrix}\psi_1\\ \psi_2\end{pmatrix} + \begin{pmatrix}U\psi_1\\ U\psi_2\end{pmatrix} + \frac{\gamma}{2\beta}(\vec{\mathrm{B}}\cdot\vec{\sigma})\begin{pmatrix}\psi_1\\ \psi_2\end{pmatrix}. \tag{A.5}$$

The last term in equation (A.5) is

$$\left(\mathrm{B}^1\sigma_1+\mathrm{B}^2\sigma_2+\mathrm{B}^3\sigma_3\right)\begin{pmatrix}\psi_1\\ \psi_2\end{pmatrix}=\begin{pmatrix}0 & \mathrm{B}^1\\ \mathrm{B}^1 & 0\end{pmatrix}\begin{pmatrix}\psi_1\\ \psi_2\end{pmatrix}+\begin{pmatrix}0 & -i\mathrm{B}^2\\ i\mathrm{B}^2 & 0\end{pmatrix}\begin{pmatrix}\psi_1\\ \psi_2\end{pmatrix}+\begin{pmatrix}\mathrm{B}^3 & 0\\ 0 & -\mathrm{B}^3\end{pmatrix}\begin{pmatrix}\psi_1\\ \psi_2\end{pmatrix}=$$

$$=\begin{pmatrix}\mathrm{B}^1\psi_2\\ \mathrm{B}^1\psi_1\end{pmatrix}+\begin{pmatrix}-i\mathrm{B}^2\psi_2\\ i\mathrm{B}^2\psi_1\end{pmatrix}+\begin{pmatrix}\mathrm{B}^3\psi_1\\ -\mathrm{B}^3\psi_2\end{pmatrix}=\begin{pmatrix}\left[\mathrm{B}^1-i\mathrm{B}^2\right]\psi_2+\mathrm{B}^3\psi_1\\ \left[\mathrm{B}^1+i\mathrm{B}^2\right]\psi_1-\mathrm{B}^3\psi_2\end{pmatrix}. \tag{A.6}$$

Given (A.6), equation (A.5) becomes

$$\begin{cases}\dfrac{i}{\beta}\partial_t\psi_1=-\alpha\beta\left(\hat{\mathrm{p}}-\dfrac{\gamma}{2\alpha\beta}\vec{\mathrm{A}}\right)^2\psi_1+U\psi_1+\dfrac{\gamma}{2\beta}\left[\left(\mathrm{B}^1-i\mathrm{B}^2\right)\psi_2+\mathrm{B}^3\psi_1\right],\\ \dfrac{i}{\beta}\partial_t\psi_2=-\alpha\beta\left(\hat{\mathrm{p}}-\dfrac{\gamma}{2\alpha\beta}\vec{\mathrm{A}}\right)^2\psi_2+U\psi_2+\dfrac{\gamma}{2\beta}\left[\left(\mathrm{B}^1+i\mathrm{B}^2\right)\psi_1-\mathrm{B}^3\psi_2\right].\end{cases} \tag{A.7}$$

Let us substitute the representation $\psi_n=|\psi_n|\exp(i\varphi_n)$ to system (A.7). We transform each term included in (A.7)

$$\partial_t\psi_n=e^{i\varphi_n}\left(\partial_t|\psi_n|+|\psi_n|i\partial_t\varphi_n\right), \tag{A.8}$$

$$\left(\hat{\mathrm{p}}-\frac{\gamma}{2\alpha\beta}\vec{\mathrm{A}}\right)^2=\hat{\mathrm{p}}^2-\frac{\gamma}{\alpha\beta}\vec{\mathrm{A}}\cdot\hat{\mathrm{p}}+\frac{\gamma^2}{4\alpha^2\beta^2}\left|\vec{\mathrm{A}}\right|^2+\frac{i\gamma\chi}{2\alpha\beta^2},$$

$$\left(\hat{\mathrm{p}}-\frac{\gamma}{2\alpha\beta}\vec{\mathrm{A}}\right)^2\psi_n=\hat{\mathrm{p}}^2\left(|\psi_n|e^{i\varphi_n}\right)-\frac{\gamma}{\alpha\beta}\vec{\mathrm{A}}\cdot\hat{\mathrm{p}}\left(|\psi_n|e^{i\varphi_n}\right)+\frac{\gamma^2}{4\alpha^2\beta^2}\left|\vec{\mathrm{A}}\right|^2|\psi_n|e^{i\varphi_n}+\frac{i\gamma\chi}{2\alpha\beta^2}|\psi_n|e^{i\varphi_n}, \tag{A.9}$$

where $\chi=\mathrm{div}_r\,\vec{\mathrm{A}}$. Let us calculate $\hat{\mathrm{p}}\psi_n$ and $\hat{\mathrm{p}}^2\psi_n$

$$\hat{\mathrm{p}}\psi_n=\frac{e^{i\varphi_n}}{\beta}\left(-i\nabla_r|\psi_n|+|\psi_n|\nabla_r\varphi_n\right), \tag{A.10}$$

$$\hat{\mathrm{p}}^2\psi_n=\frac{e^{i\varphi_n}}{\beta^2}\left(|\psi_n||\nabla_r\varphi_n|^2-\Delta_r|\psi_n|-i|\psi_n|\Delta_r\varphi_n-2i\nabla_r\varphi_n\cdot\nabla_r|\psi_n|\right). \tag{A.11}$$

Substituting expressions (A.10) and (A.11) into (A.9), we obtain

$$\begin{aligned}\left(\hat{\mathrm{p}}-\frac{\gamma}{2\alpha\beta}\vec{\mathrm{A}}\right)^2\psi_n&=\frac{e^{i\varphi_n}}{\beta^2}\left(|\psi_n||\nabla_r\varphi_n|^2-\Delta_r|\psi_n|-i|\psi_n|\Delta_r\varphi_n-2i\nabla_r\varphi_n\cdot\nabla_r|\psi_n|\right)-\\ &-\frac{\gamma e^{i\varphi_n}}{\alpha\beta^2}\left(-i\vec{\mathrm{A}}\cdot\nabla_r|\psi_n|+|\psi_n|\vec{\mathrm{A}}\cdot\nabla_r\varphi_n\right)+\frac{\gamma^2}{4\alpha^2\beta^2}\left|\vec{\mathrm{A}}\right|^2|\psi_n|e^{i\varphi_n}+\frac{i\gamma\chi}{2\alpha\beta^2}|\psi_n|e^{i\varphi_n}.\end{aligned} \tag{A.12}$$

Considering expressions (A.8) and (A.12), the first equation of system (A.7) will take the form

$$\frac{i}{\beta}e^{i\varphi_1}\left(\partial_t|\psi_1|+|\psi_1|i\partial_t\varphi_1\right)=-\frac{\alpha}{\beta}e^{i\varphi_1}\left(|\psi_1||\nabla_r\varphi_1|^2-\Delta_r|\psi_1|-i|\psi_1|\Delta_r\varphi_1-2i\nabla_r\varphi_1\cdot\nabla_r|\psi_1|\right)+$$

$$+\frac{\gamma}{\beta}e^{i\varphi_1}\left(-i\vec{A}\cdot\nabla_r|\psi_1|+|\psi_1|\vec{A}\cdot\nabla_r\varphi_1\right)-\frac{\gamma^2}{4\alpha\beta}\left|\vec{A}\right|^2|\psi_1|e^{i\varphi_1}-\frac{i\gamma\chi}{2\beta}|\psi_1|e^{i\varphi_1}+U|\psi_1|e^{i\varphi_1}+$$
$$+\frac{\gamma}{2\beta}\left[\left(B^1-iB^2\right)|\psi_2|e^{i\varphi_2}+B^3|\psi_1|e^{i\varphi_1}\right],$$

$$\begin{gathered}\frac{i}{\beta}\partial_t|\psi_1|-|\psi_1|\frac{1}{\beta}\partial_t\varphi_1=-\frac{\alpha}{\beta}|\psi_1||\nabla_r\varphi_1|^2+\frac{\alpha}{\beta}\Delta_r|\psi_1|+\frac{\gamma}{\beta}|\psi_1|\vec{A}\cdot\nabla_r\varphi_1-\frac{\gamma^2}{4\alpha\beta}\left|\vec{A}\right|^2|\psi_1|+U|\psi_1|+\\ +\frac{\gamma B_1}{2\beta}|\psi_2|\cos(\varphi_2-\varphi_1)+\frac{\gamma B_2}{2\beta}|\psi_2|\sin(\varphi_2-\varphi_1)+\frac{\gamma B_3}{2\beta}|\psi_1|+i\frac{\alpha}{\beta}|\psi_1|\Delta_r\varphi_1+i\frac{\alpha}{\beta}2\nabla_r\varphi_1\cdot\nabla_r|\psi_1|-\\ -i\frac{\gamma}{\beta}\vec{A}\cdot\nabla_r|\psi_1|-i\frac{\gamma\chi}{2\beta}|\psi_1|+i\frac{\gamma B_1}{2\beta}|\psi_2|\sin(\varphi_2-\varphi_1)-i\frac{\gamma B_2}{2\beta}|\psi_2|\cos(\varphi_2-\varphi_1).\end{gathered}\tag{A.13}$$

Equation (A.13) is complex, so it splits into a real and an imaginary part. For the real part of (A.13), we get the equation

$$-\frac{1}{\beta}\partial_t\varphi_1=-\frac{1}{4\alpha\beta}\left(4\alpha^2|\nabla_r\varphi_1|^2-4\alpha\gamma\vec{A}\cdot\nabla_r\varphi_1+\gamma^2\left|\vec{A}\right|^2\right)+U+\frac{\alpha}{\beta}\frac{\Delta_r|\psi_1|}{|\psi_1|}+$$
$$+\frac{\gamma}{2\beta}\frac{|\psi_2|}{|\psi_1|}\left[B^1\cos(\varphi_2-\varphi_1)+B^2\sin(\varphi_2-\varphi_1)\right]+\frac{\gamma B^3}{2\beta},$$
$$\begin{gathered}-\frac{1}{\beta}\partial_t\varphi_1=-\frac{1}{4\alpha\beta}\left|-\alpha\nabla_r\Phi_1+\gamma\vec{A}\right|^2+U+Q_1+\\ +\frac{\gamma}{2\beta}\left[\sqrt{\frac{\kappa_2}{\kappa_1}}\left(B^1\cos\varphi_{21}+B^2\sin\varphi_{21}\right)+B^3\right].\end{gathered}\tag{A.14}$$

Let us write the dot product for the potential $P_n$

$$\left(\vec{\mu}_\pm,\vec{B}\right)_n=\frac{\gamma}{2\beta}\left[\sqrt{\kappa_n^{-1}-1}\left(B^1\cos\varphi_{21}+B^2\sin\varphi_{21}\right)\pm B^3\right].\tag{A.15}$$

Given the notations (1.1), (1.2), and (A.15), equation (A.14) becomes equation (1.6) when $n=1$. Similarly, for the imaginary part of (A.13), we have

$$|\psi_1|\partial_t|\psi_1|=\alpha|\psi_1|^2\Delta_r\varphi_1+2\alpha\nabla_r\varphi_1\cdot|\psi_1|\nabla_r|\psi_1|-\gamma\vec{A}\cdot|\psi_1|\nabla_r|\psi_1|-\frac{\gamma\chi}{2}|\psi_1|^2+$$
$$+\frac{\gamma}{2}B^1|\psi_1||\psi_2|\sin\varphi_{21}-\frac{\gamma}{2}B^2|\psi_1||\psi_2|\cos\varphi_{21},$$
$$\partial_t|\psi_1|^2=2\alpha|\psi_1|^2\Delta_r\varphi_1+2\alpha\nabla_r\varphi_1\cdot\nabla_r|\psi_1|^2-\gamma\vec{A}\cdot\nabla_r|\psi_1|^2-\gamma|\psi_1|^2\operatorname{div}_r\vec{A}+$$
$$+\gamma|\psi_1||\psi_2|\left(B^1\sin\varphi_{21}-B^2\cos\varphi_{21}\right),$$
$$\partial_t|\psi_1|^2=\alpha\operatorname{div}_r\left(|\psi_1|^2\nabla_r\Phi_1\right)-\gamma\operatorname{div}_r\left(|\psi_1|^2\vec{A}\right)+\gamma|\psi_1||\psi_2|\left(B^1\sin\varphi_{21}-B^2\cos\varphi_{21}\right),$$
$$\partial_t f_1+\operatorname{div}_r\left[f_1\left(-\alpha\nabla_r\Phi_1+\gamma\vec{A}\right)\right]=\gamma f\sqrt{\kappa_1\kappa_2}\left(B^1\sin\varphi_{21}-B^2\cos\varphi_{21}\right).\tag{A.16}$$

Let us use the definition of the dot product (1.4)-(1.5)

$$2\beta\vec{\mu}_a(\vec{\varphi}) = \gamma(-\sin\varphi_{21} \;\; \cos\varphi_{21} \;\; 0), \;\; f_n = f\kappa_n, \tag{A.17}$$
$$2\beta f_n\left(\vec{\mu}_a(\vec{\varphi}), \vec{\mathrm{B}}\right)_n = \gamma f\kappa_n\sqrt{\kappa_n^{-1}-1}\left(\mathrm{B}^2\cos\varphi_{21} - \mathrm{B}^1\sin\varphi_{21}\right), \quad \kappa_n\sqrt{\kappa_n^{-1}-1} = \sqrt{\kappa_n(1-\kappa_n)}.$$

Given (A.17), equation (A.16) is the same as equation (1.9) for $n=1$. Given the expressions (A.8) and (A.12), the second equation of the system (A.7) will take a form

$$\frac{i}{\beta}\partial_t|\psi_2| - \frac{1}{\beta}|\psi_2|\partial_t\varphi_2 = -\frac{\alpha}{\beta}|\psi_2||\nabla_r\varphi_2|^2 + \frac{\alpha}{\beta}\Delta_r|\psi_2| + \frac{\gamma}{\beta}|\psi_2|\vec{\mathrm{A}}\cdot\nabla_r\varphi_2 - \frac{\gamma^2}{4\alpha\beta}\left|\vec{\mathrm{A}}\right|^2|\psi_2| + U|\psi_2| +$$
$$+\frac{\alpha}{\beta}i|\psi_2|\Delta_r\varphi_2 + \frac{\alpha}{\beta}2i\nabla_r\varphi_2\cdot\nabla_r|\psi_2| - i\frac{\gamma}{\beta}\vec{\mathrm{A}}\cdot\nabla_r|\psi_2| - \frac{i\gamma\chi}{2\beta}|\psi_2| + \frac{\gamma}{2\beta}\left[\left(\mathrm{B}^1 + i\mathrm{B}^2\right)|\psi_1|e^{-i\varphi_{21}} - \mathrm{B}^3|\psi_2|\right],$$

$$\frac{i}{\beta}\partial_t|\psi_2| - \frac{1}{\beta}|\psi_2|\partial_t\varphi_2 = -\frac{\alpha}{\beta}|\psi_2||\nabla_r\varphi_2|^2 + \frac{\alpha}{\beta}\Delta_r|\psi_2| + \frac{\gamma}{\beta}|\psi_2|\vec{\mathrm{A}}\cdot\nabla_r\varphi_2 - \frac{\gamma^2}{4\alpha\beta}\left|\vec{\mathrm{A}}\right|^2|\psi_2| + U|\psi_2| +$$
$$+\frac{\gamma}{2\beta}\mathrm{B}^1|\psi_1|\cos\varphi_{21} + \frac{\gamma}{2\beta}\mathrm{B}^2|\psi_1|\sin\varphi_{21} - \frac{\gamma}{2\beta}\mathrm{B}^3|\psi_2| + \frac{\alpha}{\beta}i|\psi_2|\Delta_r\varphi_2 + \tag{A.18}$$
$$+\frac{\alpha}{\beta}2i\nabla_r\varphi_2\cdot\nabla_r|\psi_2| - i\frac{\gamma}{\beta}\vec{\mathrm{A}}\cdot\nabla_r|\psi_2| - \frac{i\gamma\chi}{2\beta}|\psi_2| + i\frac{\gamma}{2\beta}\mathrm{B}^2|\psi_1|\cos\varphi_{21} - i\frac{\gamma}{2\beta}\mathrm{B}^1|\psi_1|\sin\varphi_{21}.$$

For the real part of equation (A.18), we have

$$-\frac{1}{\beta}\partial_t\varphi_2 = -\frac{1}{4\alpha\beta}\left(|\alpha\nabla_r\Phi_2|^2 - 2\alpha\gamma\vec{\mathrm{A}}\cdot\nabla_r\Phi_2 + \gamma^2\left|\vec{\mathrm{A}}\right|^2\right) + U + \mathrm{Q}_2 +$$
$$+\frac{\gamma}{2\beta}\left[\sqrt{\frac{\kappa_1}{\kappa_2}}\left(\mathrm{B}^1\cos\varphi_{21} + \mathrm{B}^2\sin\varphi_{21}\right) - \mathrm{B}^3\right]. \tag{A.19}$$

Equation (A.19), taking into account (A.15), coincides with equation (1.6) when $n=2$. For the imaginary part of (A.18), we obtain the equation

$$\partial_t|\psi_2|^2 = 2\alpha|\psi_2|^2\Delta_r\varphi_2 + 2\alpha\nabla_r\varphi_2\cdot\nabla_r|\psi_2|^2 - \gamma\vec{\mathrm{A}}\cdot\nabla_r|\psi_2|^2 - \gamma|\psi_2|^2\operatorname{div}_r\vec{\mathrm{A}} +$$
$$+\gamma|\psi_1||\psi_2|\left(\mathrm{B}^2\cos\varphi_{21} - \mathrm{B}^1\sin\varphi_{21}\right),$$
$$\partial_t f_2 + f_2\operatorname{div}_r\left(-\alpha\nabla_r\Phi_2 + \gamma\vec{\mathrm{A}}\right) + \left(-\alpha\nabla_r\Phi_2 + \gamma\vec{\mathrm{A}}\right)\cdot\nabla_r f_2 = \gamma f\sqrt{\kappa_1\kappa_2}\left(\mathrm{B}^2\cos\varphi_{21} - \mathrm{B}^1\sin\varphi_{21}\right),$$
$$\partial_t f_2 + \operatorname{div}_r\left(f_2\langle\vec{v}_2\rangle\right) = \gamma f\sqrt{\kappa_1\kappa_2}\left(\mathrm{B}^2\cos\varphi_{21} - \mathrm{B}^1\sin\varphi_{21}\right). \tag{A.20}$$

Given (A.17), equation (A.20) is the same as (1.9) when $n=2$. Adding equations (A.16) and (A.20) gives

$$\partial_t(f_1 + f_2) + \operatorname{div}_r\left(f_1\langle\vec{v}_1\rangle + f_2\langle\vec{v}_2\rangle\right) = 0 \Rightarrow \partial_t f + \operatorname{div}_r\left[f\left(\kappa_1\langle\vec{v}_1\rangle + \kappa_2\langle\vec{v}_2\rangle\right)\right] = 0, \tag{A.21}$$

where according to (1.2) $\kappa_n = f_n/f$. Taking into account the decomposition (1.1), equation (A.21) reduces to the original continuity equation. Theorem 1 is proved.

***Proof of Theorem 2.***

According to Lemma 1, the scalar potential $\Phi$ can be decomposed into $\Phi_n$ (1.2), therefore

$$2\varphi+2\pi j=2\sum_{n=1}^{2}\kappa_n\left(\varphi_n+\pi k\right)\Rightarrow \varphi+\pi j=\sum_{n=1}^{2}\kappa_n\varphi_n+\pi k\sum_{n=1}^{2}\kappa_n=\sum_{n=1}^{2}\kappa_n\varphi_n+\pi k,$$

$$\varphi=\sum_{n=1}^{2}\kappa_n\varphi_n+\pi l. \tag{A.22}$$

Expression (A.22) is the same as (1.13). The validity of expression (1.14) for the vector potential $\vec{\mathrm{A}}_\Psi=\vec{\mathrm{A}}+\vec{\mathrm{A}}_Q$ directly follows from expression (1.3), given that $\nabla_r\kappa_1=-\nabla_r\kappa_2$.

We obtain expressions for the quantum potential (1.17). From the definition $\mathrm{Q}_\Psi$ follows that

$$\frac{\beta}{\alpha}\mathrm{Q}_S=\frac{\Delta_r\sqrt{|\psi_1|^2+|\psi_2|^2}}{\sqrt{|\psi_1|^2+|\psi_2|^2}}=\frac{1}{\sqrt{|\psi_1|^2+|\psi_2|^2}}\operatorname{div}_r\frac{|\psi_1|\nabla_r|\psi_1|+|\psi_2|\nabla_r|\psi_2|}{\sqrt{|\psi_1|^2+|\psi_2|^2}}=$$

$$=\frac{\operatorname{div}_r\left(\sqrt{\kappa_1}\nabla_r|\psi_1|+\sqrt{\kappa_2}\nabla_r|\psi_2|\right)}{\sqrt{|\psi_1|^2+|\psi_2|^2}}=\frac{\sqrt{\kappa_1}\Delta_r|\psi_1|+\sqrt{\kappa_2}\Delta_r|\psi_2|}{\sqrt{|\psi_1|^2+|\psi_2|^2}}+\frac{\nabla_r\sqrt{\kappa_1}\cdot\nabla_r|\psi_1|+\nabla_r\sqrt{\kappa_2}\cdot\nabla_r|\psi_2|}{\sqrt{|\psi_1|^2+|\psi_2|^2}},$$

$$\frac{\beta}{\alpha}\mathrm{Q}_S=\kappa_1\frac{\Delta_r|\psi_1|}{|\psi_1|}+\kappa_2\frac{\Delta_r|\psi_2|}{|\psi_2|}+\frac{\nabla_r\sqrt{\kappa_1}\cdot\nabla_r|\psi_1|}{\sqrt{|\psi_1|^2+|\psi_2|^2}}+\frac{\nabla_r\sqrt{\kappa_2}\cdot\nabla_r|\psi_2|}{\sqrt{|\psi_1|^2+|\psi_2|^2}}. \tag{A.23}$$

Calculate the values $\nabla_r\sqrt{\kappa_n}$

$$\nabla_r\sqrt{\kappa_n}=\nabla_r\frac{|\psi_n|}{\sqrt{|\psi_1|^2+|\psi_2|^2}}=\frac{\sqrt{|\psi_1|^2+|\psi_2|^2}\nabla_r|\psi_n|-|\psi_n|\nabla_r\sqrt{|\psi_1|^2+|\psi_2|^2}}{|\psi_1|^2+|\psi_2|^2}=$$

$$=\frac{|\psi_1|^2\nabla_r|\psi_n|-|\psi_n||\psi_1|\nabla_r|\psi_1|+|\psi_2|^2\nabla_r|\psi_n|-|\psi_n||\psi_2|\nabla_r|\psi_2|}{\left(|\psi_1|^2+|\psi_2|^2\right)^{3/2}},$$

hence

$$\nabla_r\sqrt{\kappa_1}=|\psi_2|\frac{|\psi_2|\nabla_r|\psi_1|-|\psi_1|\nabla_r|\psi_2|}{\left(|\psi_1|^2+|\psi_2|^2\right)^{3/2}},\ \nabla_r\sqrt{\kappa_2}=|\psi_1|\frac{|\psi_1|\nabla_r|\psi_2|-|\psi_2|\nabla_r|\psi_1|}{\left(|\psi_1|^2+|\psi_2|^2\right)^{3/2}}. \tag{A.24}$$

Substituting (A.24) into (A.23) and using definition (1.7), we obtain

$$\frac{\beta}{\alpha}\mathrm{Q}_S=\kappa_1\frac{\beta}{\alpha}\mathrm{Q}_1+\kappa_2\frac{\beta}{\alpha}\mathrm{Q}_2+\frac{|\psi_2|^2\nabla_r|\psi_1|\cdot\nabla_r|\psi_1|-|\psi_1||\psi_2|\nabla_r|\psi_2|\cdot\nabla_r|\psi_1|}{\left(|\psi_1|^2+|\psi_2|^2\right)^2}+$$

$$+\frac{|\psi_1|^2\nabla_r|\psi_2|\cdot\nabla_r|\psi_2|-|\psi_1||\psi_2|\nabla_r|\psi_1|\cdot\nabla_r|\psi_2|}{\left(|\psi_1|^2+|\psi_2|^2\right)^2}=\frac{\beta}{\alpha}\mathrm{Q}+\kappa_2\frac{\nabla_r|\psi_1|\cdot\nabla_r|\psi_1|}{|\psi_1|^2+|\psi_2|^2}+\kappa_1\frac{\nabla_r|\psi_2|\cdot\nabla_r|\psi_2|}{|\psi_1|^2+|\psi_2|^2}-$$

$$-\sqrt{\kappa_1\kappa_2}\frac{\nabla_r|\psi_2|\cdot\nabla_r|\psi_1|}{|\psi_1|^2+|\psi_2|^2}-\sqrt{\kappa_1\kappa_2}\frac{\nabla_r|\psi_1|\cdot\nabla_r|\psi_2|}{|\psi_1|^2+|\psi_2|^2}=\frac{\beta}{\alpha}\mathrm{Q}+$$

$$+\frac{1}{f}\left(\kappa_2\left|\nabla_r|\psi_1|\right|^2-2\sqrt{\kappa_1\kappa_2}\nabla_r|\psi_2|\cdot\nabla_r|\psi_1|+\kappa_1\left|\nabla_r|\psi_2|\right|^2\right)=\frac{\beta}{\alpha}\mathrm{Q}+\frac{1}{f}\left(\sqrt{\kappa_2}\nabla_r|\psi_1|-\sqrt{\kappa_1}\nabla_r|\psi_2|\right)^2=$$

$$=\frac{\beta}{\alpha}\mathrm{Q}+\left(\sqrt{\kappa_1\kappa_2}\frac{\nabla_r|\psi_1|}{|\psi_1|}-\sqrt{\kappa_1\kappa_2}\frac{\nabla_r|\psi_2|}{|\psi_2|}\right)^2=\frac{\beta}{\alpha}\mathrm{Q}+\kappa_1\kappa_2\left(\nabla_r\ln|\psi_1|-\nabla_r\ln|\psi_2|\right)^2,$$

$$\mathrm{Q}_S=\mathrm{Q}+\kappa_1\kappa_2\frac{\alpha}{\beta}\left|\nabla_r\ln|\psi_1/\psi_2|\right|^2. \tag{A.25}$$

Taking into account the relation $|\psi_1/\psi_2|^2=\kappa_1/\kappa_2$ and definition (1.16), the expression (A.25) goes to (1.17). Let us derive the expression for the potential energy included in the Hamilton-Jacobi equation (i.5)

$$\mathrm{V}_\Psi=-\frac{1}{\beta}\partial_t\varphi+\frac{1}{4\alpha\beta}\left|\langle\vec{v}\rangle\right|^2. \tag{A.26}$$

Using (A.22) and the expansion (1.1), we will calculate the right-hand side of (A.26). Let us start with the first term on the right-hand side of (A.26).

$$\frac{1}{\beta}\partial_t\varphi=\kappa_1\frac{1}{\beta}\partial_t\varphi_1+\kappa_2\frac{1}{\beta}\partial_t\varphi_2+\frac{\varphi_1}{\beta}\partial_t\kappa_1+\frac{\varphi_2}{\beta}\partial_t\kappa_2=-\kappa_1\mathrm{H}_1-\kappa_2\mathrm{H}_2-\frac{2\kappa_1\kappa_2}{\beta}\varphi_{21}\partial_t\ln|\psi_1/\psi_2|, \tag{A.27}$$

where it is taken into account that

$$\partial_t\kappa_1=\partial_t\frac{|\psi_1|^2}{|\psi_1|^2+|\psi_2|^2}=2|\psi_1||\psi_2|\frac{|\psi_2|\partial_t|\psi_1|-|\psi_1|\partial_t|\psi_2|}{\left(|\psi_1|^2+|\psi_2|^2\right)^2}=2\kappa_1\kappa_2\frac{\partial_t|\psi_1|}{|\psi_1|}-2\kappa_1\kappa_2\frac{\partial_t|\psi_2|}{|\psi_2|}=$$
$$=2\kappa_1\kappa_2\left(\partial_t\ln|\psi_1|-\partial_t\ln|\psi_2|\right)=2\kappa_1\kappa_2\partial_t\ln|\psi_1/\psi_2|,\quad \partial_t\kappa_2=-\partial_t\kappa_1. \tag{A.28}$$

Using the definitions for $\mathrm{H}_n$ (1.6) and $\mathcal{V}$ (1.15), the expression (A.27) takes the form

$$\frac{1}{\beta}\partial_t\varphi=\frac{1}{4\alpha\beta}\left(\kappa_1\left|\langle\vec{v}_1\rangle\right|^2+\kappa_2\left|\langle\vec{v}_2\rangle\right|^2\right)-\mathcal{V}-\frac{2\kappa_1\kappa_2}{\beta}\varphi_{21}\partial_t\ln|\psi_1/\psi_2|. \tag{A.29}$$

Substitute (A.29) into the original expression (A.26)

$$\mathrm{V}_\Psi=\mathcal{V}+\frac{1}{4\alpha\beta}\left(\left|\kappa_1\langle\vec{v}_1\rangle+\kappa_2\langle\vec{v}_2\rangle\right|^2-\kappa_1\left|\langle\vec{v}_1\rangle\right|^2-\kappa_2\left|\langle\vec{v}_2\rangle\right|^2\right)+\frac{2\kappa_1\kappa_2}{\beta}\varphi_{21}\partial_t\ln|\psi_1/\psi_2|. \tag{A.30}$$

Let us transform the second term on the right-hand side of expression (A.30)

$$\left|\kappa_1\langle\vec{v}_1\rangle+\kappa_2\langle\vec{v}_2\rangle\right|^2-\kappa_1\left|\langle\vec{v}_1\rangle\right|^2-\kappa_2\left|\langle\vec{v}_2\rangle\right|^2=-\kappa_1\kappa_2\left(\left|\langle\vec{v}_1\rangle\right|^2+\left|\langle\vec{v}_2\rangle\right|^2-2\langle\vec{v}_1\rangle\cdot\langle\vec{v}_2\rangle\right)=$$
$$=-\kappa_1\kappa_2\left|\langle\vec{v}_1\rangle-\langle\vec{v}_2\rangle\right|^2=-\kappa_1\kappa_2\left|-\alpha\nabla_r\Phi_1+\alpha\nabla_r\Phi_2\right|^2=-4\alpha^2\kappa_1\kappa_2\left|\nabla_r\varphi_{21}\right|^2, \tag{A.31}$$

where we used $\kappa_1^2 = \kappa_1 - \kappa_1\kappa_2$ and $\kappa_2^2 = \kappa_2 - \kappa_1\kappa_2$. Substitute (A.31) into (A.30), we will obtain

$$\mathrm{V}_\Psi = \mathcal{V} - \frac{1}{\beta}\kappa_1\kappa_2\left(\alpha\left|\nabla_r\varphi_{21}\right|^2 - \varphi_{21}\partial_t \ln\left|\psi_1/\psi_2\right|^2\right). \tag{A.32}$$

Expression (A.32) proves the validity of representations (1.14) and (1.16). The validity of the left-hand side of expression (1.15) directly follows from the definitions $\mathrm{Q}$, $\mathrm{P}$ (1.15) and $\mathcal{V}_n$ (1.6)

$$\mathrm{V} = \kappa_1\mathrm{V}_1 + \kappa_2\mathrm{V}_2 = \kappa_1\left(U + \mathrm{Q}_1 + \mathrm{P}_1\right) + \kappa_2\left(U + \mathrm{Q}_2 + \mathrm{P}_2\right) = U + \mathrm{Q} + \mathrm{P}. \tag{A.33}$$

Theorem 2 has been proved.

***Proof of Theorem 3.***

Let us differentiate equation (1.6) and use the expansion (1.1) from Lemma 1

$$-\frac{1}{2\alpha\beta}\partial_t\alpha\nabla_r\Phi_n = -\frac{1}{4\alpha\beta}\nabla_r\left|\langle\vec{v}_n\rangle\right|^2 + \nabla_r\mathcal{V}_n = -\frac{\gamma}{2\alpha\beta}\langle\vec{v}_n\rangle\times\vec{\mathrm{B}} - \frac{1}{2\alpha\beta}\left(\langle\vec{v}_n\rangle\cdot\nabla_r\right)\langle\vec{v}_n\rangle + \nabla_r\mathcal{V}_n,$$

$$\frac{1}{2\alpha\beta}\partial_t\left(-\alpha\nabla_r\Phi_n + \gamma\vec{\mathrm{A}}\right) - \frac{\gamma}{2\alpha\beta}\partial_t\vec{\mathrm{A}} + \frac{1}{2\alpha\beta}\left(\langle\vec{v}_n\rangle\cdot\nabla_r\right)\langle\vec{v}_n\rangle = \nabla_r\mathcal{V}_n - \frac{\gamma}{2\alpha\beta}\langle\vec{v}_n\rangle\times\vec{\mathrm{B}},$$

$$\left(\partial_t + \langle\vec{v}_n\rangle\cdot\nabla_r\right)\langle\vec{v}_n\rangle = -\gamma\left(-\partial_t\vec{\mathrm{A}} - \frac{2\alpha\beta}{\gamma}\nabla_r\mathcal{V}_n + \langle\vec{v}_n\rangle\times\vec{\mathrm{B}}\right), \tag{A.34}$$

where it is taken into account

$$\langle\vec{v}_n\rangle\times\left(\nabla_r\times\langle\vec{v}_n\rangle\right) = \gamma\langle\vec{v}_n\rangle\times\vec{\mathrm{B}} = \frac{1}{2}\nabla_r\left|\langle\vec{v}_n\rangle\right|^2 - \left(\langle\vec{v}_n\rangle\cdot\nabla_r\right)\langle\vec{v}_n\rangle, \tag{A.35}$$

$$\mathrm{curl}_r\langle\vec{v}_n\rangle = \gamma\,\mathrm{curl}_r\vec{\mathrm{A}} = \gamma\vec{\mathrm{B}}.$$

Equation (A.34) is the same as equation (2.1). We obtain the relations between electric fields $\vec{\mathrm{E}}$ и $\vec{\mathrm{E}}_\Psi$. It directly follows from expressions (i.8) and (1.17)-(1.18) that

$$\vec{\mathrm{E}}_\Psi = \vec{\mathrm{E}}_S + \vec{\mathrm{E}}_Q = -\partial_t\vec{\mathrm{A}} - \frac{2\alpha\beta}{\gamma}\nabla_r U_S - \partial_t\vec{\mathrm{A}}_Q - \frac{2\alpha\beta}{\gamma}\nabla_r\mathrm{Q}_S = -\partial_t\vec{\mathrm{A}} - \partial_t\vec{\mathrm{A}}_Q -$$

$$-\frac{2\alpha\beta}{\gamma}\nabla_r\left[U + \mathrm{P} + \frac{\kappa_1\kappa_2}{\beta}\left(\varphi_{21}\partial_t K - \frac{\alpha}{4}\left|\nabla_r K\right|^2 - \alpha\left|\nabla_r\varphi_{21}\right|^2\right) + \mathrm{Q} + \frac{\alpha}{4\beta}\kappa_1\kappa_2\left|\nabla_r K\right|^2\right] =$$

$$= \vec{\mathrm{E}} - \frac{2\alpha\beta}{\gamma}\nabla_r\mathrm{P} - \frac{2\alpha\beta}{\gamma}\nabla_r\mathrm{Q} - \frac{2\alpha}{\gamma}\nabla_r\left[\kappa_1\kappa_2\left(\varphi_{21}\partial_t K - \frac{\alpha}{4}\left|\nabla_r K\right|^2 - \alpha\left|\nabla_r\varphi_{21}\right|^2\right)\right] - \partial_t\vec{\mathrm{A}}_Q -$$

$$-\frac{\alpha^2}{2\gamma}\nabla_r\left(\kappa_1\kappa_2\left|\nabla_r K\right|^2\right) = \vec{\mathrm{E}} - \kappa_1\frac{2\alpha\beta}{\gamma}\nabla_r\mathrm{P}_1 - \frac{2\alpha\beta}{\gamma}\kappa_2\nabla_r\mathrm{P}_2 - \frac{2\alpha\beta}{\gamma}\kappa_1\nabla_r\mathrm{Q}_1 - \frac{2\alpha\beta}{\gamma}\kappa_2\nabla_r\mathrm{Q}_2 -$$

$$-\frac{2\alpha\beta}{\gamma}\mathrm{P}_1\nabla_r\kappa_1 - \frac{2\alpha\beta}{\gamma}\mathrm{P}_2\nabla_r\kappa_2 - \frac{2\alpha\beta}{\gamma}\mathrm{Q}_1\nabla_r\kappa_1 - \frac{2\alpha\beta}{\gamma}\mathrm{Q}_2\nabla_r\kappa_2 + \frac{2\alpha}{\gamma}\partial_t\left(\varphi_{21}\nabla_r\kappa_1\right) -$$

$$-\frac{2\alpha}{\gamma}\nabla_r\left[\kappa_1\kappa_2\left(\varphi_{21}\partial_t K - \alpha\left|\nabla_r\varphi_{21}\right|^2\right)\right],$$

$$\vec{\mathrm{E}}_{\Psi}=\vec{\mathrm{E}}+\kappa_1\vec{\mathrm{E}}_{\mathrm{P}_1}+\kappa_1\vec{\mathrm{E}}_{\mathrm{Q}_1}+\kappa_2\vec{\mathrm{E}}_{\mathrm{P}_2}+\kappa_2\vec{\mathrm{E}}_{\mathrm{Q}_2}+\frac{2\alpha\beta}{\gamma}(\mathrm{P}_{21}+\mathrm{Q}_{21})\nabla_r\kappa_1+ \tag{A.36}$$

$$+\frac{2\alpha}{\gamma}\partial_t(\varphi_{21}\nabla_r\kappa_1)-\frac{2\alpha}{\gamma}\nabla_r(\kappa_1\kappa_2\varphi_{21}\partial_t K)+\frac{2\alpha^2}{\gamma}\nabla_r\left(\kappa_1\kappa_2|\nabla_r\varphi_{21}|^2\right).$$

Using the definition (2.4)

$$\vec{\mathrm{E}}_{\psi}=\kappa_1\vec{\mathrm{E}}_{\psi_1}+\kappa_2\vec{\mathrm{E}}_{\psi_2}=\vec{\mathrm{E}}+\kappa_1\left(\vec{\mathrm{E}}_{\mathrm{Q}_1}+\vec{\mathrm{E}}_{\mathrm{P}_1}\right)+\kappa_2\left(\vec{\mathrm{E}}_{\mathrm{Q}_2}+\vec{\mathrm{E}}_{\mathrm{P}_2}\right), \tag{A.37}$$

the expression for energy $\mathcal{V}_{21}$ (1.12) and (1.16), for representation (A.36) we obtain

$$\vec{\mathrm{E}}_{\Psi}=\vec{\mathrm{E}}_{\psi}+\frac{2\alpha}{\gamma}\partial_t(\varphi_{21}\nabla_r\kappa_1)-\frac{2\alpha}{\gamma}\partial_t\varphi_{21}\nabla_r\kappa_1-\frac{2\alpha}{\gamma}\nabla_r(\kappa_1\kappa_2\varphi_{21}\partial_t K)+\frac{1}{2\gamma}\left(\left|\langle\vec{v}_2\rangle\right|^2-\left|\langle\vec{v}_1\rangle\right|^2\right)\nabla_r\kappa_1+$$

$$+\frac{2\alpha^2}{\gamma}\nabla_r\left(\kappa_1\kappa_2|\nabla_r\varphi_{21}|^2\right)=\vec{\mathrm{E}}_{\psi}+\frac{1}{\gamma}(\alpha\Phi_{21}\partial_t\nabla_r\kappa_1)-\frac{1}{\gamma}\nabla_r\left[\alpha\Phi_{21}(\kappa_2+\kappa_1)\partial_t\kappa_1\right]+$$

$$+\frac{1}{2\gamma}\left(\left|\langle\vec{v}_2\rangle\right|^2-\left|\langle\vec{v}_1\rangle\right|^2\right)\nabla_r\kappa_1+\frac{1}{2\gamma}\nabla_r\left(\kappa_1\kappa_2|\alpha\nabla_r\Phi_{21}|^2\right),$$

$$\vec{\mathrm{E}}_{\Psi}=\vec{\mathrm{E}}_{\psi}+\delta\vec{\mathrm{E}}_{\psi}, \tag{A.38}$$

$$\gamma\delta\vec{\mathrm{E}}_{\psi}=\partial_t\kappa_1(-\alpha\nabla_r\Phi_{21})+\frac{1}{2}\left[(-\alpha\nabla_r\Phi_{21})\cdot(\langle\vec{v}_1\rangle+\langle\vec{v}_2\rangle)\right]\nabla_r\kappa_1+$$

$$+\frac{1}{2}|\alpha\nabla_r\Phi_{21}|^2\nabla_r(\kappa_1\kappa_2)+\kappa_1\kappa_2(-\alpha\nabla_r\Phi_{21}\cdot\nabla_r)(-\alpha\nabla_r\Phi_{21}). \tag{A.39}$$

Let us transform the remaining term $\delta\vec{\mathrm{E}}_{\psi}$ (A.39). We will start from the second and the third term in (A.39)

$$\frac{1}{2}\left[(-\alpha\nabla_r\Phi_{21})\cdot(\langle\vec{v}_1\rangle+\langle\vec{v}_2\rangle)\right]\nabla_r\kappa_1+\frac{1}{2}|\alpha\nabla_r\Phi_{21}|^2\nabla_r(\kappa_1\kappa_2)=$$

$$=\frac{\nabla_r\kappa_1}{2}\left[(-\alpha\nabla_r\Phi_{21})\cdot(\langle\vec{v}_1\rangle+\langle\vec{v}_2\rangle)+|\alpha\nabla_r\Phi_{21}|^2(\kappa_2-\kappa_1)\right]=$$

$$=\frac{\nabla_r\kappa_1}{2}\left[(-\alpha\nabla_r\Phi_{21})\cdot\left(\langle\vec{v}_1\rangle+\langle\vec{v}_2\rangle-\alpha\nabla_r\Phi_{21}(\kappa_2-\kappa_1)\right)\right]. \tag{A.40}$$

Transform the expression in parentheses (A.40)

$$\langle\vec{v}_1\rangle+\langle\vec{v}_2\rangle-\alpha\nabla_r\Phi_{21}(\kappa_2-\kappa_1)=-\alpha\kappa_1\nabla_r\Phi_1-\alpha\nabla_r\Phi_1+2\gamma\vec{\mathrm{A}}-\alpha\nabla_r\Phi_2+$$

$$+(-\alpha\nabla_r\Phi_2+\alpha\nabla_r\Phi_1)(\kappa_2-\kappa_1)=2\gamma\vec{\mathrm{A}}-\alpha\nabla_r\Phi_1+\alpha(1-\kappa_1)\nabla_r\Phi_1-\alpha\kappa_1\nabla_r\Phi_1-\alpha\nabla_r\Phi_2-$$

$$-\alpha\kappa_2\nabla_r\Phi_2+\alpha(1-\kappa_2)\nabla_r\Phi_2=2\gamma\vec{\mathrm{A}}-2\kappa_1\alpha\nabla_r\Phi_1-2\kappa_2\alpha\nabla_r\Phi_2=$$

$$=2\left(-\kappa_1\alpha\nabla_r\Phi_1+\kappa_1\gamma\vec{\mathrm{A}}-\kappa_2\alpha\nabla_r\Phi_2+\kappa_2\gamma\vec{\mathrm{A}}\right),$$

$$\langle\vec{v}_1\rangle+\langle\vec{v}_2\rangle-\alpha\nabla_r\Phi_{21}(\kappa_2-\kappa_1)=2\langle\vec{v}\rangle. \tag{A.41}$$

Substituting (A.41) into (A.40), we obtain

$$\frac{1}{2}\Big[\big(-\alpha\nabla_r\Phi_{21}\big)\cdot\big(\langle\vec{v}_1\rangle+\langle\vec{v}_2\rangle\big)\Big]\nabla_r\kappa_1+\frac{1}{2}\left|\alpha\nabla_r\Phi_{21}\right|^2\nabla_r\left(\kappa_1\kappa_2\right)=-\alpha\big(\nabla_r\Phi_{21}\cdot\langle\vec{v}\rangle\big)\nabla_r\kappa_1. \quad (A.42)$$

Given (A.42), the expression (A.39) becomes

$$\begin{aligned}\gamma\delta\vec{E}_\psi &=\Big[\partial_t\kappa_1+\kappa_1\kappa_2\left(-\alpha\nabla_r\Phi_{21}\cdot\nabla_r\right)\Big]\left(-\alpha\nabla_r\Phi_{21}\right)+\nabla_r\kappa_1\Big[\left(-\alpha\nabla_r\Phi_{21}\right)\cdot\langle\vec{v}\rangle\Big]=\\ &=\Big[\partial_t\kappa_1+\big(\langle\vec{v}\rangle\cdot\nabla_r\kappa_1\big)-\big(\langle\vec{v}\rangle\cdot\nabla_r\kappa_1\big)+\kappa_1\kappa_2\left(-\alpha\nabla_r\Phi_{21}\cdot\nabla_r\right)\Big]\left(-\alpha\nabla_r\Phi_{21}\right)+\\ &+\Big[\left(-\alpha\nabla_r\Phi_{21}\right)\cdot\langle\vec{v}\rangle\Big]\nabla_r\kappa_1=\left[\frac{d\kappa_1}{dt}+\kappa_1\kappa_2\left(-\alpha\nabla_r\Phi_{21}\cdot\nabla_r\right)\right]\left(-\alpha\nabla_r\Phi_{21}\right)+\\ &+\Big[\left(-\alpha\nabla_r\Phi_{21}\right)\cdot\langle\vec{v}\rangle\Big]\nabla_r\kappa_1-\big(\langle\vec{v}\rangle\cdot\nabla_r\kappa_1\big)\left(-\alpha\nabla_r\Phi_{21}\right),\end{aligned}$$

$$\gamma\delta\vec{E}_\psi=\left[\frac{d\kappa_1}{dt}+\kappa_1\kappa_2\left(-\alpha\nabla_r\Phi_{21}\cdot\nabla_r\right)\right]\left(-\alpha\nabla_r\Phi_{21}\right)-\gamma\langle\vec{v}\rangle\times\vec{B}_Q, \quad (A.43)$$

where we use

$$\nabla_r\kappa_1\Big[\left(-\alpha\nabla_r\Phi_{21}\right)\cdot\langle\vec{v}\rangle\Big]-\big(\langle\vec{v}\rangle\cdot\nabla_r\kappa_1\big)\left(-\alpha\nabla_r\Phi_{21}\right)=-\alpha\langle\vec{v}\rangle\times\left(\nabla_r\kappa_1\times\nabla_r\Phi_{21}\right)=-\gamma\langle\vec{v}\rangle\times\vec{B}_Q, \quad (A.44)$$

since according to (1.3)

$$\vec{B}_Q=-\frac{\alpha}{\gamma}\left(\nabla_r\kappa_1\times\nabla_r\Phi_1+\nabla_r\kappa_2\times\nabla_r\Phi_2\right)=\frac{\alpha}{\gamma}\nabla_r\kappa_1\times\nabla_r\Phi_{21}. \quad (A.45)$$

Substitution of the additional term (A.43) into (A.39) proves the validity of (2.5). Theorem 3 is proved.

***Proof of Theorem 4***

Since the fields $\vec{E},\vec{B}$ satisfy Maxwell's equations, hence the following equations are valid

$$\partial_t\vec{D}+\vec{J}=\mathrm{curl}_r\vec{H},\qquad \mathrm{div}_r\vec{D}=\rho,\qquad \mathrm{div}_r\vec{B}=0,\qquad -\partial_t\vec{B}=\mathrm{curl}_r\vec{E}, \quad (A.46)$$

where the current density $\vec{J}$, in general, it may differ from the current density $\vec{J}_\Psi$. The difference between them is defined as

$$\vec{J}_\Psi=\vec{J}+\vec{J}_Q\Rightarrow\vec{J}=\vec{J}_\Psi-\vec{J}_Q. \quad (A.47)$$

Let us write the relation between fields $\vec{E}$ и $\vec{E}_\Psi$. According to (i.11) and (1.14)-(1.15) expression for $\vec{E}_\Psi$ takes the form

$$\vec{E}_\Psi=-\partial_t\vec{A}-\partial_t\vec{A}_Q-\frac{1}{q}\nabla_r\left(U+Q+P+\delta\mathcal{V}\right)=\vec{E}-\partial_t\vec{A}_Q-\frac{1}{q}\nabla_r\left(Q+P+\delta\mathcal{V}\right),$$

$$\vec{E}=\vec{E}_\Psi+\partial_t\vec{A}_Q+\frac{1}{q}\nabla_r\left(Q+P+\delta\mathcal{V}\right). \quad (A.48)$$

Similarly for the magnetic field

$$\vec{\mathrm{B}}_\Psi = \vec{\mathrm{B}}_S + \vec{\mathrm{B}}_Q = \vec{\mathrm{B}} + \vec{\mathrm{B}}_Q,\ \ \vec{\mathrm{B}} = \vec{\mathrm{B}}_\Psi - \vec{\mathrm{B}}_Q = \mu_0\left(\vec{\mathrm{H}}_\Psi - \vec{\mathrm{H}}_Q\right). \tag{A.49}$$

Substitute (A.48)-(A.49) into the first equation of (A.49)

$$c\partial_0\,\vec{\mathrm{D}}_\Psi + \varepsilon_0 c^2 \partial_0\partial^0\,\vec{\mathrm{A}}_Q + \varepsilon_0\frac{c}{q}\partial_0\nabla_r\left(\mathrm{Q}+\mathrm{P}+\delta\mathcal{V}\right) + \vec{J}_\Psi - \vec{J}_Q = \mathrm{curl}_r\,\vec{\mathrm{H}}_\Psi - \mathrm{curl}_r\,\vec{\mathrm{H}}_Q,$$

$$\partial_t\,\vec{\mathrm{D}}_\Psi + \vec{\mathrm{J}}_\Psi = \mathrm{curl}_r\,\vec{\mathrm{H}}_\Psi + \vec{\delta}, \tag{A.50}$$

$$\vec{\delta} = -\mathrm{curl}_r\,\vec{\mathrm{H}}_Q - \varepsilon_0 c^2\partial_0\partial^0\,\vec{\mathrm{A}}_Q - \varepsilon_0\frac{c}{q}\partial_0\nabla_r\left(\mathrm{Q}+\mathrm{P}+\delta\mathcal{V}\right) + \vec{\mathrm{J}}_Q. \tag{A.51}$$

Transform the expression $\delta$ (A.51)

$$\begin{aligned}\mu_0\delta &= -\nabla_r\,\mathrm{div}_r\,\vec{\mathrm{A}}_Q + \Delta_r\,\vec{\mathrm{A}}_Q - \partial_0\partial^0\,\vec{\mathrm{A}}_Q - \frac{1}{qc}\partial_0\nabla_r\left(\mathrm{Q}+\mathrm{P}+\delta\mathcal{V}\right) + \mu_0\,\vec{\mathrm{J}}_Q = \\ &= -\Box\vec{\mathrm{A}}_Q + \mu_0\,\vec{\mathrm{J}}_Q - \nabla_r\left[\mathrm{div}_r\,\vec{\mathrm{A}}_Q + \frac{1}{qc}\partial_0\left(\mathrm{Q}+\mathrm{P}+\delta\mathcal{V}\right)\right],\end{aligned} \tag{A.52}$$

where $\mathrm{curl}_r\,\mathrm{curl}_r\,\vec{\mathrm{A}}_Q = \nabla_r\,\mathrm{div}_r\,\vec{\mathrm{A}}_Q - \Delta_r\,\vec{\mathrm{A}}_Q$.

According to the theorem conditions the fields $\vec{\mathrm{E}}_\Psi$, $\vec{\mathrm{B}}_\Psi$ satisfy the Lorenz $\Psi$ - gauge condition, hence

$$\mathrm{div}_r\,\vec{\mathrm{A}} + \mathrm{div}_r\,\vec{\mathrm{A}}_Q + \frac{1}{qc}\partial_0\left(U + \mathrm{Q}+\mathrm{P}+\delta\mathcal{V}\right) = 0,$$

$$\mathrm{div}_r\,\vec{\mathrm{A}}_Q = -\frac{1}{qc}\partial_0\left(\mathrm{Q}+\mathrm{P}+\delta\mathcal{V}\right), \tag{A.53}$$

where the usual Lorenz gauge is used. Substituting (A.53) into (A.52)

$$\mu_0\vec{\delta} = -\Box\vec{\mathrm{A}}_Q + \mu_0\,\vec{\mathrm{J}}_Q, \tag{A.54}$$

from the equation above, taking using (2.10), we obtain

$$\Box\vec{\mathrm{A}}_Q = \mu_0\,\vec{\mathrm{J}}_Q \Rightarrow \vec{\delta} = \vec{\theta} \Rightarrow \partial_t\,\vec{\mathrm{D}}_\Psi + \vec{\mathrm{J}}_\Psi = \mathrm{curl}_r\,\vec{\mathrm{H}}_\Psi. \tag{A.55}$$

By definition (2.8), the charge density is $\rho = \mathrm{div}_r\,\vec{\mathrm{D}}$, $\rho_\Psi = \mathrm{div}_r\,\vec{\mathrm{D}}_\Psi$. Let us calculate their values. On the one hand, from expression (A.48), it follows that

$$\rho = \rho_\Psi + \varepsilon_0\partial_t\,\mathrm{div}_r\,\vec{\mathrm{A}}_Q + \frac{1}{q}\varepsilon_0\Delta_r\left(\mathrm{Q}+\mathrm{P}+\delta\mathcal{V}\right) = \rho_\Psi - \varepsilon_0\frac{1}{q}\Box\left(\mathrm{V}_\Psi - U\right),$$

$$\rho = \rho_\Psi - \varepsilon_0\Box\left(\varphi_\Psi - \varphi\right), \tag{A.56}$$

on the other hand, considering the expression for $\vec{\mathrm{E}}_\Psi$ (i.11) and the Lorenz $\Psi$ - gauge condition, we obtain

$$\rho_\Psi = \varepsilon_0 \operatorname{div}_r \vec{\mathrm{E}}_\Psi = -\varepsilon_0 \partial_t \operatorname{div}_r \vec{\mathrm{A}}_\Psi - \frac{\varepsilon_0}{q} \Delta_r \mathrm{V}_\Psi = \frac{\varepsilon_0}{q} \Box \mathrm{V}_\Psi = \varepsilon_0 \Box \varphi_\Psi. \tag{A.57}$$

Substituting (A.57) into (A.56), we go to the equation

$$\rho = \varepsilon_0 \Box \varphi. \tag{A.58}$$

Equations (A.57) and (A.58) are the same as equations (2.15). Note that an important relation follows from (1.14)-(1.15) and (1.18), which we will need later.

$$\mathrm{V}_\Psi - U = Q + P + \delta\mathcal{V} = \mathrm{Q}_S + \delta U. \tag{A.59}$$

Given (1.18), we transform the expression for $\vec{\mathrm{E}}_S$ (i.11), we get

$$\vec{\mathrm{E}}_S = -\partial_t \vec{\mathrm{A}} - \frac{1}{q} \nabla_r U_S = -\partial_t \vec{\mathrm{A}} - \nabla_r \varphi - \nabla_r \delta\varphi = \vec{\mathrm{E}} - \nabla_r \delta\varphi, \tag{A.60}$$

which is the same as $\vec{\mathrm{E}}_S$ from (2.17). Using expressions (A.53) and (A.59), let us calculate $\rho_Q$ and $\rho_S$. Start from definition (2.8), we get

$$\frac{\rho_Q}{\varepsilon_0} \overset{\text{def}}{=} \varepsilon_0 \operatorname{div}_r \vec{\mathrm{E}}_Q = -\partial_t \operatorname{div}_r \vec{\mathrm{A}}_Q - \Delta_r \varphi_Q = \frac{1}{q} \partial_0 \partial^0 \left(\mathrm{Q} + \mathrm{P} + \delta\mathcal{V}\right) - \Delta_r \varphi_Q =$$

$$= \frac{1}{q} \partial_0 \partial^0 \left(\mathrm{Q}_S + \delta U\right) - \Delta_r \varphi_Q = \partial_0 \partial^0 \left(\varphi_Q + \delta\varphi\right) - \Delta_r \varphi_Q,$$

$$\rho_Q = \varepsilon_0 \Box \varphi_Q + \varepsilon_0 \partial_0 \partial^0 \delta\varphi, \tag{A.61}$$

$$\frac{\rho_S}{\varepsilon_0} \overset{\text{def}}{=} \operatorname{div}_r \vec{\mathrm{E}}_S = \operatorname{div}_r \vec{\mathrm{E}} - \Delta_r \delta\varphi = \frac{\rho}{\varepsilon_0} - \Delta_r \delta\varphi,$$

$$\rho_S = \rho + \delta\rho. \tag{A.62}$$

Expression (A.61) proves the validity of (2.16) for $\rho_Q$. Adding expressions (A.61) and (A.62), we get

$$\rho_\Psi = \operatorname{div}_r \vec{\mathrm{D}}_\Psi = \operatorname{div}_r \vec{\mathrm{D}}_Q + \operatorname{div}_r \vec{\mathrm{D}}_S = \rho + \delta\rho + \rho_Q. \tag{A.63}$$

Expressions (A.62) and (A.63) prove the validity of (2.14). From (A.63) and (A.62), we obtain an expression for $\rho_S$ (2.16)

$$\rho_S = \rho_\Psi - \rho_Q = \varepsilon_0 \Box\left(\varphi_\Psi - \varphi_Q\right) - \varepsilon_0 \partial_0 \partial^0 \delta\varphi = \varepsilon_0 \Box \varphi_S - \varepsilon_0 \partial_0 \partial^0 \delta\varphi, \tag{A.64}$$

where we used the definition $\mathrm{V}_\Psi$ (i.6), that is $\mathrm{V}_\Psi - \mathrm{Q}_S = U_S$ and therefore $\varphi_\Psi - \varphi_Q = \varphi_S$ (compare with (2.13)). From (A.62), (A.64) and (A.58) we obtain

$$\rho_S = \varepsilon_0 \Box \varphi_S - \varepsilon_0 \partial_0 \partial^0 \delta\varphi = \rho + \delta\rho = \varepsilon_0 \Box \varphi - \varepsilon_0 \Delta_r \delta\varphi,$$

$$\Box \varphi_S = \Box \varphi + \partial_0 \partial^0 \delta\varphi - \Delta_r \delta\varphi = \Box\left(\varphi + \delta\varphi\right) \Rightarrow \varphi_S = \varphi + \delta\varphi, \tag{A.65}$$

which is valid up to a certain function $b$: $\Box b=0$. Expression (A.65) corresponds to (2.13). Substituting (A.59) in (A.48), we obtain $\vec{\mathrm{E}}_\Psi$ from (2.17).

Based on the definition of $\vec{\mathrm{B}}_\Psi=\mathrm{curl}_r\,\vec{\mathrm{A}}_\Psi$ follows that $\mathrm{div}_r\,\vec{\mathrm{B}}_\Psi=0$, similarly from the definition $\vec{\mathrm{E}}_\Psi=-\partial_t\,\vec{\mathrm{A}}_\Psi-\frac{1}{q}\nabla_r\,\mathrm{V}_\Psi$ follows that $-\partial_t\,\vec{\mathrm{B}}_\Psi=\mathrm{curl}_r\,\vec{\mathrm{E}}_\Psi$. These two equations, together with equations (A.55) and (A.63), form the Maxwell system of equations for fields $\vec{\mathrm{E}}_\Psi$ и $\vec{\mathrm{B}}_\Psi$. Theorem 4 is proved.

## Appendix B

***Proof of Corollary 1.***

Expressions (3.1) and (3.2) are the standard result of the Pauli equation solution for quantum systems with a uniform field $\vec{\mathrm{B}}=\mathrm{curl}_r\,\vec{\mathrm{A}}=(b_2-b_1)\vec{e}_z$. Let us calculate the values of the expansion coefficients $\kappa_n$

$$\kappa_n=\frac{|\psi_n|^2}{|\psi|^2}=\frac{|\Upsilon|^2|c_n|^2}{|\Upsilon|^2}=|c_n|^2=const. \tag{B.1}$$

From (B.1) it follows those coefficients $\kappa_n$ are constant values, and the condition $\kappa_1+\kappa_2=1$ (1.2) is automatically satisfied by the normalization condition (3.1). Let us write the consequences of theorems 1-4 based on condition (B.1). From Lemma 1, the validity of (3.6) follows. From Theorem 1 (Remark 1), it follows that

$$\left(\vec{\mu}_a(\vec{\varphi})\cdot\vec{\mathrm{B}}\right)_n=0\;\Rightarrow\;\mathrm{P}_n=-(-1)^n\mu_B\,\mathrm{B}\;\Rightarrow\;\partial_t f_n+\mathrm{div}_r\,\vec{J}_n=0, \tag{B.2}$$

which proves the validity of (3.3) and the continuity equation (3.9). From Theorem 2

$$\psi_n=|\Upsilon||c_n|\exp\left\{i\left[\varphi_\Upsilon+\varphi_{c_n}+(-1)^n\omega_0 t\right]\right\}\Rightarrow\varphi_n=\varphi_\Upsilon+\varphi_{c_n}+(-1)^n\omega_0 t, \tag{B.3}$$

$$\varphi_{21}=\varphi_\Upsilon+\varphi_{c_2}+\omega_0 t-\varphi_\Upsilon-\varphi_{c_1}+\omega_0 t=2\omega_c t+\varphi_{c_2}-\varphi_{c_1}$$

$$\varphi=\kappa_1\left(\varphi_\Upsilon+\varphi_{c_1}-\omega_0 t\right)+\kappa_2\left(\varphi_\Upsilon+\varphi_{c_2}+\omega_0 t\right)+\pi l=\varphi_\Upsilon+\omega_0 t(\kappa_2-\kappa_1)+\kappa_1\varphi_{c_1}+\kappa_2\varphi_{c_2}+\pi l, \tag{B.4}$$

where $\Psi(\vec{r},t)=\pm|\Upsilon|\exp(i\varphi)$. Expressions (B.3) and (B.4) are the same as (3.8). From (B.1) and (1.14)-(1.17) we obtain

$$\delta\mathrm{Q}=0\;\Rightarrow\;\mathrm{Q}_S=\mathrm{Q},\quad \mathrm{Q}_n=-\frac{\hbar^2}{2m}\frac{\Delta_r|\psi_n|}{|\psi_n|}=-\frac{\hbar^2}{2m}\frac{\Delta_r|\Upsilon|}{|\Upsilon|}=-\frac{\hbar^2}{2m}\frac{\Delta_r|\Psi|}{|\Psi|}=\mathrm{Q}_S, \tag{B.5}$$

$$\delta\mathcal{V}=0\;\Rightarrow\;\mathrm{V}_\Psi=U_S+\mathrm{Q}_S=\mathcal{V},\qquad \delta U=\mathrm{P}\;\Rightarrow\;U_S=U+(\kappa_1-\kappa_2)\mu_B\,\mathrm{B}, \tag{B.6}$$

$$\mathcal{V}_n=U+\mathrm{Q}_n-(-1)^n\mu_B\,\mathrm{B}=U_S+\mathrm{Q}_S-\left[(-1)^n+\kappa_1-\kappa_2\right]\mu_B\,\mathrm{B}, \tag{B.7}$$

$$\vec{\mathrm{E}}_{\psi_n}=-\frac{1}{q}\nabla_r\,\mathrm{V}_\Psi=\vec{\mathrm{E}}+\vec{\mathrm{E}}_\mathrm{Q}=\vec{\mathrm{E}}_\Psi,\;\vec{\mathrm{E}}=-\frac{1}{q}\nabla_r U=-\frac{1}{q}\nabla_r U_S=\vec{\mathrm{E}}_S,\;\vec{\mathrm{E}}_{\mathrm{Q}_n}=\vec{\mathrm{E}}_\mathrm{Q},\;\vec{\mathrm{E}}_{\mathrm{P}_n}=0. \tag{B.8}$$

Expressions (B.5)-(B.8) prove correctness (3.4)-(3.5) and (3.7). Substitution of $\varphi_\Upsilon = -\frac{\mathcal{E}}{\hbar} t$ into (B.3)-(B.4), and further into (1.1) gives a relation of velocities $\langle \vec{v}_n \rangle = \frac{e}{m} \vec{\mathrm{A}} = \langle \vec{v} \rangle$. As a result the Hamilton-Jacobi equations (1.6) take the form of (3.11). Sum of equations (3.11) gives us (3.10), corresponding to (i.5). Corollary 1 is proved.

***Proof of Theorem 5***

Solution of the equation (3.2) we will look for solutions based on the $\Psi$ -model [13] with phase $\varphi_\Upsilon$

$$\varphi_\Upsilon(r,t) = \ell\phi - \frac{\mathcal{E}}{\hbar} t + \pi l,\ \ell, l \in \mathbb{Z} \Rightarrow \langle \vec{v} \rangle = \frac{\hbar}{m} \nabla_r \varphi_\Upsilon + \frac{e}{m} \vec{\mathrm{A}} = \left( \frac{\hbar \ell}{m\rho} + \omega_0 \rho \right) \vec{e}_\phi, \tag{B.9}$$

and $\phi$ is the azimuthal angle and $\rho = r \sin\theta = x^2 + y^2$. From (B.9) and (3.8) and (1.1)-(1.2) we obtain (3.17). Let us notice that phase $\varphi_\Upsilon$ is defined up to an additive term $\pi l$. The value $e^{i\pi l}$ determines the sign preceding the wave function $\Upsilon = |\Upsilon| \exp(i\varphi_\Upsilon)$, therefore, we use the representation

$$\Upsilon(\rho,\phi,z,t) = R(\rho) Z(z) \exp\left( i\ell\phi - i\frac{\mathcal{E}}{\hbar} t \right) = R(\rho) Z(z) \exp(i\bar{\varphi}_\Upsilon), \tag{B.10}$$

where $R(\rho), Z(z) \in \mathbb{R}$. Let us perform a sequence of transformations

$$i\vec{\mathrm{A}} \cdot \nabla_r \Upsilon = i|\vec{\mathrm{A}}| \vec{e}_\phi \cdot \nabla_r \left( R(\rho) Z(z) e^{i\bar{\varphi}_\Upsilon} \right) = i|\vec{\mathrm{A}}| \vec{e}_\phi \cdot \nabla_r \left( RZe^{i\bar{\varphi}_\Upsilon} \right) = i\frac{1}{2} \mathrm{B} \rho RZ \frac{1}{\rho} \partial_\phi e^{i\bar{\varphi}_\Upsilon},$$

$$i\hbar \frac{e}{m} \vec{\mathrm{A}} \cdot \nabla_r \Upsilon = -\frac{e\mathrm{B}\hbar}{2m} \ell RZe^{i\bar{\varphi}_\Upsilon} = -\hbar\omega_0 \ell RZe^{i\bar{\varphi}_\Upsilon}, \tag{B.11}$$

$$\frac{1}{2m}\left(\hat{\mathrm{p}} + e\vec{\mathrm{A}}\right)^2 \Upsilon = \frac{\hat{\mathrm{p}}^2}{2m} \Upsilon + \frac{e}{m} \vec{\mathrm{A}} \cdot \hat{\mathrm{p}} \Upsilon + \frac{e^2}{2m} |\vec{\mathrm{A}}|^2 \Upsilon = -\frac{\hbar^2}{2m} \Delta_r \Upsilon + \frac{e^2 \mathrm{B}^2}{8m} \rho^2 \Upsilon - i\hbar \frac{e}{m} \vec{\mathrm{A}} \cdot \nabla_r \Upsilon =$$

$$= -\frac{\hbar^2}{2m} \left( R''Z + \frac{1}{\rho} R'Z - \frac{\ell^2}{\rho^2} RZ + RZ'' \right) e^{i\bar{\varphi}_\Upsilon} + \frac{m\omega_0^2 \rho^2}{2} RZe^{i\bar{\varphi}_\Upsilon} + \hbar\omega_0 \ell RZe^{i\bar{\varphi}_\Upsilon},$$

$$\frac{1}{2m}\left(\hat{\mathrm{p}} + e\vec{\mathrm{A}}\right)^2 \Upsilon = \left[ -\frac{\hbar^2}{2m} \left( \frac{R''}{R} + \frac{1}{\rho} \frac{R'}{R} - \frac{\ell^2}{\rho^2} + \frac{Z''}{Z} \right) + \frac{m\omega_0^2 \rho^2}{2} + \hbar\omega_0 \ell \right] RZe^{i\bar{\varphi}_\Upsilon}. \tag{B.12}$$

Substituting (B.12) into (3.2), we get

$$-\frac{\hbar^2}{2m} \left( \frac{R''}{R} + \frac{1}{\rho} \frac{R'}{R} - \frac{\ell^2}{\rho^2} \right) + U - \mathcal{E} + \frac{m\omega_0^2 \rho^2}{2} + \hbar\omega_0 \ell - \frac{\hbar^2}{2m} \frac{Z''}{Z} = 0,$$

$$-\frac{\hbar^2}{2m} \left( \frac{R''}{R} + \frac{1}{\rho} \frac{R'}{R} - \frac{\ell^2}{\rho^2} \right) + \frac{m\Omega^2 \rho^2}{2} - \mathcal{E} + \hbar\omega_0 \ell = \frac{\hbar^2}{2m} \frac{Z''}{Z} - \frac{m\Omega^2 z^2}{2} = \lambda = const. \tag{B.13}$$

For axial part of $Z(z)$ from (B.13) follows, that the limited solution has the form

$$Z(z)=e^{-\frac{\xi^2 z^2}{2}}H_l(\xi z),\ \lambda_l=-\hbar\Omega\left(l+\frac{1}{2}\right). \tag{B.14}$$

Let us transform the equation for the radial part (B.13) and take into account (B.14)

$$R''+\frac{1}{\rho}R'+\left[\frac{2m\mathcal{E}}{\hbar^2}-\frac{2m\omega_0}{\hbar}\ell-\xi^2(2l+1)-\xi^4\rho^2-\frac{\ell^2}{\rho^2}\right]R=0. \tag{B.15}$$

Making a replacement $R(\rho)=\rho^{|\ell|}e^{-\frac{1}{2}\xi^2\rho^2}F(\rho^2)$ and calculating derivatives

$$R'=e^{-\frac{1}{2}\xi^2\rho^2}\left(|\ell|\rho^{|\ell|-1}F-\xi^2\rho^{|\ell|+1}F+2\rho^{|\ell|+1}F'\right), \tag{B.16}$$

$$e^{\frac{1}{2}\xi^2\rho^2}R''=-\xi^2|\ell|\rho^{|\ell|}F+|\ell|(|\ell|-1)\rho^{|\ell|-2}F+\xi^4\rho^{|\ell|+2}F-\xi^2(|\ell|+1)\rho^{|\ell|}F-$$
$$-2\xi^2\rho^{|\ell|+2}F'+2|\ell|\rho^{|\ell|}F'-2\xi^2\rho^{|\ell|+2}F'+2(|\ell|+1)\rho^{|\ell|}F'+4\rho^{|\ell|+2}F'',$$
$$e^{\frac{1}{2}\xi^2\rho^2}\rho^{-|\ell|}R''=4\rho^2F''+\left(4|\ell|+2-4\xi^2\rho^2\right)F'+\left(\frac{\ell^2-|\ell|}{\rho^2}+\xi^4\rho^2-\xi^2-2\xi^2|\ell|\right)F. \tag{B.17}$$

Substituting (B.16) and (B.17) into (B.15), we obtain

$$4\rho^2F''+4\left(|\ell|+1-\xi^2\rho^2\right)F'+$$
$$+\left(\frac{\ell^2-|\ell|}{\rho^2}-\frac{\ell^2}{\rho^2}+\frac{|\ell|}{\rho^2}+\xi^4\rho^2-\xi^4\rho^2-\xi^2-\xi^2-\xi^2(2l+1)-2\xi^2|\ell|+\frac{2m\mathcal{E}}{\hbar^2}-\frac{2m\omega_0}{\hbar}\ell\right)F=0.$$

$$\rho^2F''+\left(|\ell|+1-\xi^2\rho^2\right)F'+\mu F=0, \tag{B.18}$$

$$\frac{m\mathcal{E}}{2\hbar^2}-\frac{m\omega_0}{2\hbar}\ell-\frac{2l+3}{4}\xi^2-\frac{1}{2}\xi^2|\ell|=\mu>0. \tag{B.19}$$

Making a replacement $\bar{\rho}=\xi^2\rho^2$, $F(\rho^2)=G(\bar{\rho})$, consequently, $F'=\xi^2G'$ и $F''=\xi^4G''$ . Thus, equation (B.18) takes the known form

$$\bar{\rho}G''+\left(|\ell|+1-\bar{\rho}\right)G'+sG=0,\ \mu=s\xi^2\ \Rightarrow\ F(\rho^2)=L_s^{(|\ell|)}\left(\xi^2\rho^2\right). \tag{B.20}$$

Using the expression for $\mu$ (B.20), we find the energy levels $\mathcal{E}$ (B.19)

$$\mathcal{E}_{s,l,\ell}=\hbar\omega_0\ell+\hbar\Omega\left(2s+|\ell|+l+\frac{3}{2}\right) \tag{B.21}$$

Now we calculate the normalization constant $N^2_{s,\ell,l}$

$$\int\limits_{\mathbb{R}^3} |\psi_{s,\ell}|^2 d^3r = N_{s,\ell,l}^2 \int\limits_0^{2\pi} d\phi \int\limits_{-\infty}^{+\infty} H_l^2(\xi z) e^{-\xi^2 z^2} dz \int\limits_0^{+\infty} \rho^{2|\ell|} e^{-\xi^2\rho^2} L_s^{(|\ell|)}(\xi^2\rho^2) L_s^{(|\ell|)}(\xi^2\rho^2)\rho d\rho =$$
$$= N_{s,\ell,l}^2 2^l l! \frac{\pi^{3/2}}{\xi^{2|\ell|+3}} \int\limits_0^{+\infty} x^{|\ell|} e^{-x} L_s^{(|\ell|)}(x) L_s^{(|\ell|)}(x) dx = N_{s,\ell,l}^2 \frac{2^l l! \pi^{3/2}}{\xi^{2|\ell|+3} s!} \Gamma(s+|\ell|+1) = 1, \tag{B.22}$$

where $x = \xi^2\rho^2$. Expression (3.16) for a quantum potential is obtained from (3.4) and from the Hamilton-Jacobi equation (3.10). Let us write the Hamilton-Jacobi equation for the wave function $\Upsilon$. Using (B.9) and (B.21)

$$\mathcal{E}_{s,\ell,l} = \frac{m}{2}|\langle\vec{v}\rangle|^2 + U + \mathrm{Q}_\Upsilon = \frac{\hbar^2\ell^2}{2m\rho^2} + \hbar\omega_0\ell + \frac{m\omega_0^2\rho^2}{2} + \frac{m\,\omega^2\rho^2}{2} + \frac{m\Omega^2 z^2}{2} + \mathrm{Q}_\Upsilon,$$
$$\mathrm{Q}_\Upsilon = \hbar\Omega\left(|\ell| + l + 2s + \frac{3}{2}\right) - \frac{\hbar^2\ell^2}{2m\rho^2} - \frac{m\Omega^2 r^2}{2}. \tag{B.23}$$

Since $|\Upsilon| = |\Psi| = |\psi|$, that $\mathrm{Q}_\Upsilon = \mathrm{Q}_S = \mathrm{Q}_n = \mathrm{Q}$. Theorem 5 has been proved.

***Calculation of the magnetic momentum (3.25)***

$$\vec{M}_{s,\ell,l} = -\frac{e}{2}\int\limits_{-\infty}^{+\infty} dz \int\limits_0^{2\pi} d\phi \int\limits_0^{+\infty} f_{s,\ell,l}\left(\vec{r}\times\langle\vec{v}\rangle_\ell\right)\rho d\rho =$$
$$= -\frac{e}{2}\int\limits_{-\infty}^{+\infty} dz \int\limits_0^{2\pi} d\phi \int\limits_0^{+\infty}\left(\frac{\hbar\ell}{m\rho} + \omega_0\rho\right) f_{s,\ell,l}\left(\rho\vec{e}_\rho\times\vec{e}_\phi + z\vec{e}_z\times\vec{e}_\phi\right)\rho d\rho =$$
$$= -\vec{e}_z\frac{e}{2}\int\limits_{-\infty}^{+\infty} dz \int\limits_0^{2\pi} d\phi \int\limits_0^{+\infty}\left(\frac{\hbar\ell}{m\rho} + \omega_0\rho\right) f_{s,\ell,l}\rho^2 d\rho + \frac{e}{2}\int\limits_{-\infty}^{+\infty} zdz \int\limits_0^{2\pi} d\phi \int\limits_0^{+\infty}\left(\frac{\hbar\ell}{m\rho} + \omega_0\rho\right) f_{s,\ell,l}\vec{e}_\rho\rho d\rho =$$
$$= -\vec{e}_z\frac{e\hbar\ell}{2m}\int\limits_{-\infty}^{+\infty} dz \int\limits_0^{2\pi} d\phi \int\limits_0^{+\infty} f_{s,\ell,l}\rho d\rho - \vec{e}_z\frac{e\omega_0}{2}\int\limits_{-\infty}^{+\infty} dz \int\limits_0^{2\pi} d\phi \int\limits_0^{+\infty} f_{s,\ell,l}\rho^3 d\rho +$$
$$+\frac{e\hbar\ell}{2m}\int\limits_{-\infty}^{+\infty} zdz \int\limits_0^{2\pi}\begin{pmatrix}\cos\phi\\ \sin\phi\\ 0\end{pmatrix} d\phi \int\limits_0^{+\infty} f_{s,\ell,l}(\rho,z)d\rho + \frac{e\omega_0}{2}\int\limits_{-\infty}^{+\infty} zdz \int\limits_0^{2\pi}\begin{pmatrix}\cos\phi\\ \sin\phi\\ 0\end{pmatrix} d\phi \int\limits_0^{+\infty} f_{s,\ell,l}(\rho,z)\rho^2 d\rho,$$
$$\vec{M}_{s,\ell,l} = -\vec{e}_z\mu_B\ell - \vec{e}_z\pi e\omega_0\int\limits_{-\infty}^{+\infty} dz \int\limits_0^{+\infty} f_{s,\ell,l}\rho^3 d\rho. \tag{B.24}$$

Calculation of the integral in (B.24)

$$I_{s,\ell,l} = \pi e\omega_0\int\limits_{-\infty}^{+\infty} dz \int\limits_0^{+\infty} f_{s,\ell,l}\rho^3 d\rho = \pi e\omega_0 N_{s,\ell,l}^2 \int\limits_{-\infty}^{+\infty} e^{-\xi^2 z^2} H_l^2(\xi z) dz \int\limits_0^{+\infty} \rho^{2|\ell|+3} e^{-\xi^2\rho^2}\left[L_s^{(|\ell|)}(\xi^2\rho^2)\right]^2 d\rho =$$
$$= -2^l l! N_{s,\ell,l}^2 \frac{e\omega_0\pi^{3/2}}{2\xi^{2|\ell|+5}} \int\limits_0^{+\infty} \tau^{|\ell|+1} e^{-\tau}\left[L_s^{(|\ell|)}(\tau)\right]^2 d\tau, \tag{B.25}$$

where $\tau = \xi^2\rho^2$. Using the power expansion of the Laguerre generalized polynomials in (B.25)

$$I_{s,\ell,l} = -2^l l! N_{s,\ell,l}^2 \frac{e\omega_0 \pi^{3/2}}{2\xi^{2|\ell|+5}} \sum_{k,n=0}^{s} C_{s+|\ell|}^{s-k} C_{s+|\ell|}^{s-n} \frac{(-1)^{k+n}}{n!k!} \int_0^{+\infty} \tau^{|\ell|+n+k+1} e^{-\tau} d\tau =$$

$$= -\frac{s!\xi^{2|\ell|+3}}{\pi^{3/2}\Gamma(s+|\ell|+1)} \frac{e\omega_0 \pi^{3/2}}{2\xi^{2|\ell|+5}} \sum_{k,n=0}^{s} C_{s+|\ell|}^{s-k} C_{s+|\ell|}^{s-n} \frac{(-1)^{k+n}}{n!k!} (|\ell|+n+k+1)! =$$

$$= -\frac{s!}{(s+|\ell|)!} \frac{e\hbar\omega_0}{2m\Omega} \sum_{k,n=0}^{s} C_{s+|\ell|}^{s-k} C_{s+|\ell|}^{s-n} \frac{(-1)^{k+n}}{n!k!} (|\ell|+n+k+1)! =$$

$$= -\mu_B \frac{\omega_0}{\Omega} \sum_{k,n=0}^{s} (-1)^{k+n} (n+1) C_{s+|\ell|}^{s-n} C_s^k C_{|\ell|+n+k+1}^{n+1} = -\mu_B \frac{\omega_0}{\Omega} (|\ell|+2s+1). \quad \text{(B.26)}$$

Substituting (B.26) into (B.24), we get (3.26).

**Acknowledgments**

The study was conducted under the state assignment of Lomonosov Moscow State University.